\documentclass[3p]{elsart}
\usepackage{graphicx}
\usepackage{multirow} 
\usepackage{amssymb}
\usepackage{amsmath}
\usepackage{enumerate}
\usepackage{ulem}
\usepackage{color}
\usepackage{soulutf8}
\usepackage[switch]{lineno}

\begin{document}

\begin{frontmatter}

\title{
Radio pulses from ultra-high energy \\ 
atmospheric showers as the superposition \\ 
of Askaryan and geomagnetic mechanisms \\
}

\author[USC]{Jaime Alvarez-Mu\~niz,}
\ead{jaime.alvarezmuniz@gmail.com}
\author[USC]{Washington R. Carvalho Jr.,}
\ead{carvajr@gmail.com}
\author[UH]{Harm Schoorlemmer,}
\ead{harmscho@phys.hawaii.edu}
\author[USC]{Enrique Zas}
\ead{zas@fpaxp1.usc.es}

\address[USC]{Departamento de F\'\i sica de Part\'\i culas
\& Instituto Galego de F\'\i sica de Altas Enerx\'\i as,
Universidade de Santiago de Compostela,\\ 
15782 Santiago de Compostela, Spain}

\address[UH]{University of Hawaii at Manoa, 
Department of Physics and Astronomy, Honolulu,
Hawaii 96822, USA}

\begin{abstract}
Radio emission in atmospheric showers is currently interpreted in terms
of radiation due to the deviation of the charged particles in the magnetic field of the Earth and 
to the charge excess (Askaryan effect). Each of these mechanisms
has a distinctive polarization. The complex signal patterns can be
qualitatively explained as the interference (superposition) of the fields induced
by each mechanism. In this work we explicitly and quantitatively test a simple phenomenological 
model based on this idea. The model is constructed by isolating each of the two components at 
the simulation level and by making use of approximate symmetries for each of the contributions 
separately. The results of the model are then checked against full ZHAireS Monte Carlo simulations 
of the electric field calculated from first principles.  
We show that the simple model describes radio emission at a few percent level in a wide range 
of shower-observer geometries and on a shower-by-shower basis. As a consequence, this 
approach provides a simple method to reduce the computing time needed to accurately predict 
the electric field of radio pulses emitted from air showers, with many practical applications 
in experimental situations of interest.
\end{abstract}

\end{frontmatter}

\section{Introduction}

In the last few years there has been a flurry of activity to
explore the potential of the radio technique to study ultra high
energy cosmic rays (UHECRs) \cite{Huege_ICRC13}. The technique is attractive because it
requires detectors which are relatively cheap and is not limited by a small duty cycle, unlike the
fluorescence technique \cite{Nagano-Watson}. Several experimental initiatives such as AERA \cite{AERA}
and EASIER \cite{EASIER} in coincidence with the Pierre Auger Observatory \cite{Auger}, 
CODALEMA \cite{CODALEMA}, LOPES \cite{LOPES}, TREND \cite{TREND}, LOFAR \cite{LOFAR} 
and Tunka-Rex \cite{Tunka-Rex_ICRC13}, have been exploring emission from air showers mainly in the 30-80 MHz
frequency range, studying the relation of the pulses
to composition \cite{LOPES_muons_radio,LOFAR_Xmax} and trying to understand the lateral distribution of
the emission \cite{Allan_Nature,LOPES_LDF,AERA,LOFAR}. In addition the fortuitous detection of pulses from air
showers with the ANITA balloon flown detector \cite{ANITA_UHECR} has revealed that the
pulses from these showers extend to the GHz regime, with more experimental evidence
also obtained by the CROME experiment \cite{CROME}. 
These studies are considered of primordial importance to explore 
the viability of the radio technique either to provide an alternative design
of future air shower arrays covering areas in the range of
10,000~km$^2$ and above, or to use them as complementary detectors that could help
to constrain the composition of the cosmic rays on a shower-by-shower basis. 

On the phenomenological side there has also been a lot of
progress. Calculations have been made based on the macroscopic
treatment of the current densities that arise in these showers \cite{Scholten_MGMR,Scholten_EVA}, 
and also on more detailed simulations that use the superposition principle 
to calculate the emission adding the contributions from single
particle sub-tracks \cite{REAS3,SELFAS,ZHAireS,COREAS}. The results of these calculations, that reproduce
pulses extending into the GHz regime as detected by ANITA \cite{ZHAireS_ANITA}, are now 
in reasonably good agreement \cite{Huege_ARENA12} and they 
have been successfully tested against data \cite{LOPES_COREAS,ZHAireS_ANITA,AERA,AERA_pol,Huege_ICRC13}. 
As a result there is much
more confidence in the calculations and it is
believed that it is now possible to infer properties of the shower
such as direction, energy and shower maximum using only measurements of the radio pulses in antennas 
scattered at ground level \cite{Huege_radio_mass,Konstantinov_JCAP,deVries_radio_mass,Scholten_Xmax,Huege_ICRC13}. 

Radio emission in atmospheric showers is currently interpreted in terms
of two dominant radiation mechanisms, namely the excess negative
charge predicted by Askaryan in the 1960s (the Askaryan mechanism) \cite{Askaryan}, 
and that induced by the interactions of the particles with the magnetic field of the Earth
that induces a net current perpendicular to the shower axis, the geomagnetic mechanism \cite{Kahn-Lerche}.
These two mechanisms induce distinctive electric field polarizations.
Moreover the complex signal patterns observed at ground can be qualitatively explained as the interference 
(superposition) of the fields induced by each mechanism.
In this work we quantitatively test this idea.
For that purpose we have devised a phenomenological model 
that relies on standard approximate assumptions \cite{Scholten_MGMR} 
and experimental observations of the characteristics of the radiation \cite{Allan,Prescott,CODALEMA,AERA_pol}. 
The model is constructed by isolating each of the two components at 
the simulation level and by making use of approximate symmetries for each of the contributions 
separately. The results of the model are subsequently checked against full ZHAireS Monte Carlo 
simulations of the electric field calculated from first principles.  
We show that the simple model describes radio emission at a few percent level in a wide range 
of shower-observer geometries and on a shower-by-shower basis.

The devised method reduces the calculation of the pattern at ground 
level to the pattern along a single line which can be sampled at a few
points. This has a drastic effect on the required computing resources.
This is of utmost importance since the study of the capabilities of detectors 
exploiting the radio technique as well as the reconstruction of cosmic ray shower 
parameters in detail requires full
simulations of the shower and the radio pulses at a large variety of 
locations. 
Needless to say that a comprehensive parameterization of
the patterns of the electric field at ground level would be most welcome
as an alternative \cite{LOFAR_param_paper}. The approach presented here
can be considered a step forward in this direction, since it can also simplify the
creation of parameterizations.

This paper is organized as follows.
In Section \ref{sec:basics} we briefly review the main properties of radio emission
in atmospheric cosmic ray showers. 
In Section \ref{sec:model} we describe the model to obtain the electric field induced
by ultra-high energy cosmic ray showers in the atmosphere. 
In Section \ref{sec:results} we compare the electric field calculated
with the model with that obtained in full Monte Carlo simulations.
Finally, in Section \ref{sec:conclusions} we conclude the paper and discuss
future prospects. 

\section{Basics of radio emission in air showers}
\label{sec:basics}

\subsection{Radio emission mechanisms}
\label{sec:mechanisms}
Radio emission from showers initiated by cosmic rays in the atmosphere 
is discussed in terms of two mechanisms: the deflection of charges in
the magnetic field of the Earth, the geomagnetic effect, and the development of an excess
charge as the shower entrains electrons from the atmosphere into the
shower front, the Askaryan mechanism. 
The different nature of the two
mechanisms becomes apparent when we consider shower
development without the magnetic field and only the Askaryan
contribution exists. 

The solution of Maxwell's equations in the transverse gauge relates
the vector potential to the transverse current density, which in the
limit of far observation distances is simply the projection of the
current orthogonal to the direction of observation \cite{Jackson,ZHS92,ZHS13}. This is evident 
in a microscopic approach where currents are regarded as a sample 
of small particle sub-tracks of charge $q$ moving at constant velocity 
$\vec v$, each of them contributing in proportion to $q \vec
v_\perp$~\cite{ZHS91,ZHS13}, where $v_\perp$ refers to the projection of the
sub-track into a plane perpendicular to the observation direction. 
The fact that the index of refraction is not constant in the
atmosphere, that the particle tracks are not perfectly parallel to the shower
axis and that the observation direction changes for each particle track
complicates the picture but does not alter this fact. 

The two mechanisms can be related to
different components of the current density, perpendicular and
parallel to the shower axis. The perpendicular component of the
current is related to the geomagnetic mechanism and results from the Lorentz force induced by the geomagnetic field
$\vec B$ that accelerates the charges in the direction $q \vec v \times \vec B$\footnote{This has been referred 
to as a {\sl transverse} current in
the sense that it lies in a plane perpendicular to the shower axis.}. For the
shower as a whole this current scales with  
$\vert \vec{B}\vert\sin\alpha$, with $\alpha$ the angle between $\vec{V}$
and $\vec{B}$,  where $\vec{V}\simeq<\vec v>$ is assumed to be parallel to
shower axis. So the magnitude of the vector potential,
hence the electric field, depends strongly on both the zenith,
$\theta$, and azimuth, $\phi$, angles of the shower, which determine
the value of $\alpha$. Since electrons and positrons deviate in
opposite directions they both contribute with the same sign to this component. 
The polarization to be expected for the vector
potential is simply given by the projection of the  $-\vec V \times
\vec B$ direction onto a plane transverse to the observation direction. For
observers positioned along the shower axis this coincides with 
$-\vec V \times \vec B$. This is the main mechanism for the radiation in
extensive air showers. 

The component of the current that is parallel to the shower axis 
is related to the Askaryan mechanism and arises because of the excess
electrons in the shower front. 
The magnitude of the
vector potential from this contribution is directly related to the excess charge as
the shower develops in the atmosphere and the arrival direction
of the shower has little effect on the magnitude of this excess. 
The vector potential is polarized along the projection of the
shower axis onto the plane transverse to the observation
direction, and has radial symmetry. It is thus zero along the shower axis
and increases in proportion to $\sin \beta$, where $\beta$ is the
angle between the shower axis and the observation direction \cite{ZHS92,ZHS13}. 
The polarization of the vector potential has thus a strong radial
component which increases as the observer deviates further away from
shower axis. 

\subsection{Cherenkov like effects}
\label{sec:ring}

The fact that the refractive index $n$ of the atmosphere is greater than 1 is
of utmost importance to understand the radio emission from air
showers \cite{Scholten_PRL,Scholten_EVA,ZHAireS}. Since radio waves in air travel slightly slower than the shower
particle front, shock wave effects similar to those observed in
Cherenkov radiation play an important role in the time
dependence of the electric field as seen by an observer on ground \cite{Scholten_EVA,ZHAireS}. This in
turn determines the amplitude and frequency spectrum of the observed
radiation. In a simplified one-dimensional shower model \cite{ZHAireS_ANITA}, the 
emission from the point in the shower development that
is seen by a particular observer at the Cherenkov angle arrives first
and, more importantly, the emission from a region around it arrives with a 
much smaller delay than the rest of the shower. This is the
result of the relation between observation, $t$, and retarded, $t^*$,
times which is such that $t$ has an absolute minimum with $dt/dt^*=0$ 
in the Cherenkov direction~\cite{Scholten_PRL,ZHAireS}. The vector potential has an abrupt growth
at this minimum $t$ due to an effective ``time compression'' of the
contributions from a relatively large region of the shower. This induces a strong
and narrow pulse, coherent up to frequencies in the $\sim$ GHz
and with an enhanced power \cite{ANITA_UHECR,CROME,ZHAireS_ANITA}. An observer thus becomes much more 
sensitive to the region of the shower seen at angles close to the Cherenkov angle. 
When the antenna lies in the Cherenkov direction with respect to the
maximum of the shower $X_{\rm max}$, the detected electric
fields are the largest, since more particles contribute to either of
the mechanisms. These points of maximum radiation define a circular
ring-like region on the plane perpendicular to the shower axis, or equivalently an
elliptical region when projected onto the ground given by the intersection 
of a Cherenkov cone centered at $X_{\rm max}$ with ground level. 
The major ($r_a$) and minor ($r_b$) axis of the ellipse are approximately
given by\footnote{It is worth noting that the center of the ellipse is not 
exactly at the shower core as Eq.~(\ref{eq:calcring}) suggests. This is a small effect since the
 angle (opening angle of the cone) is typically smaller than
$1^\circ$, we nevertheless use exact expressions in the model 
described in Section \ref{sec:model}.}:  

\begin{eqnarray}
\label{eq:calcring}
\nonumber
r_a&=&\left[\tan{(\theta + \theta_{\rm Cher})} - \tan{\theta}\right]~h_{X_{\rm
    max}}\\ 
\\ 
\nonumber 
r_b&=&\frac{\tan{\theta_{\rm Cher}}}{\cos{\theta}}~h_{X_{\rm max}}
\end{eqnarray}
where $\theta$ is the zenith angle of the shower, $h_{X_{\rm max}}$ is the
height above ground of the position of shower maximum, and $\theta_{\rm Cher}$ 
is the Cherenkov angle at that height\footnote{The refractive index
 decreases with increasing atmospheric altitude and $\theta_{\rm Cher}$ follows
 the same behavior.}. 
The major and minor axes of the Cherenkov ellipse on the ground scale
with $h_{X_{\rm max}}$ and also increase as the angle $\theta$ rises. 


\subsection{Radiation pattern at ground level}
\label{sec:groundpattern}
\begin{figure}
\begin{center}
\scalebox{0.7}{\includegraphics{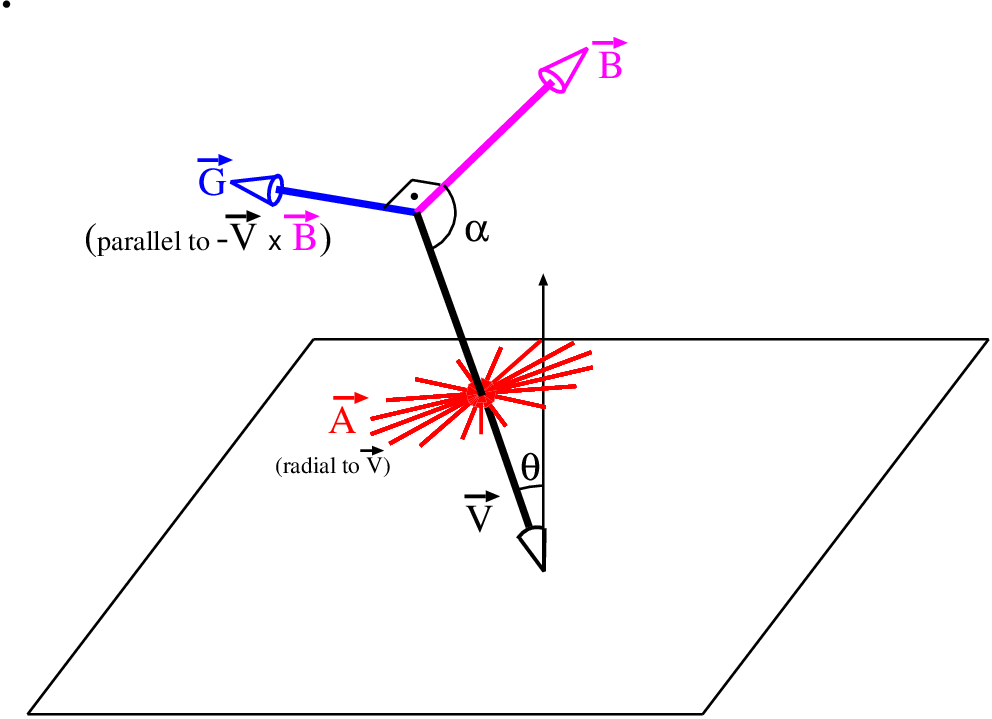}}
\caption{
Scheme of the axis of a shower  
  along with the
  polarization of the emission due to the geomagnetic ($\vec{G}$) and Askaryan
  ($\vec{A}$)  mechanisms. $\vec{B}$ is the geomagnetic field pointing North
  and $\vec{V}$ marks the direction of shower axis. The amplitudes of the geomagnetic 
  and Askaryan fields drawn in the sketch are for illustrative purposes only and not to scale.}
\label{fig:geoscheme-ask-geo-B-axis}
\end{center}
\end{figure}

The electric fields induced by the Askaryan and geomagnetic mechanisms 
interfere with each other to produce the total field at the observer position 
\cite{Scholten_MGMR,REAS3,SELFAS,ZHAireS}. 
Since the amplitude of the geomagnetic field is strongly dependent on shower
arrival direction, while the Askaryan field is not, the relative contributions of these
two mechanisms vary strongly with shower geometry. For some particular
arrival directions, Askaryan emission can actually dominate the radio emission,
e.g. when the shower axis is parallel to the direction of the magnetic field,
and thus $\alpha\simeq 0^\circ$. Some examples will be shown below. 
The region where the signal is largest is pretty much 
concentrated at the intersection of the ground plane with 
a Cherenkov cone having its vertex at the shower maximum.
This sets the scale of
the radiation pattern at ground, given by $r_a$ and $r_b$ in Eq.~(\ref{eq:calcring}). 
As the Cherenkov angle is very small, typically less than $1^\circ$, the
direction of observation of the shower region where the contribution is 
largest is nearly parallel to the shower axis and this allows some approximate
simplifications. 
Taking the direction of observation parallel to the shower direction, the Askaryan 
polarization, $\hat a$, is radial with respect to the shower axis while the
geomagnetic polarization, $\hat g$, has a constant direction given by
$-\vec V \times \vec B$ (see Fig.~\ref{fig:geoscheme-ask-geo-B-axis}). The different
patterns of the polarization vector for the two mechanisms create asymmetries 
in the footprint of the radio signal on the ground that, depending on 
the geometry of the shower and the observer position, can be very
sizeable \cite{Scholten_MGMR,REAS3,SELFAS,ZHAireS}. 
The full description of the radio emission pattern at ground level requires a two
dimensional function.

\section{A simple model}
\label{sec:model}

Here we describe   
a simple model that makes use of approximate
symmetries which are only present when the 
two dominant emission mechanisms are considered separately.
Using as input the separate
geomagnetic and Askaryan contributions to the electric field at only a few
simulated antenna positions, we are able to predict with great accuracy the
total electric field at any position on the ground. We assume that 
the amplitudes of both contributions depend only on the radial distance to
the shower axis on the shower plane, and exploit the 
approximate polarization patterns described in the previous section
for both mechanisms. This considerably reduces the CPU time needed to obtain an
accurate description of the two-dimensional pattern of the electric field on
the ground which will be
described in terms of $r$, the distance to shower core, and $\phi$,
the polar angle of the vector from the impact point of the shower 
to the observer. 

\subsection{Assumptions}
\label{sec:assumptions}

Our approach is based on a few assumptions regarding the radio emission. 
Some stem from  well-known features of the electric field induced by 
the geomagnetic and Askaryan mechanisms: 

\begin{enumerate}[(A)]

\item The polarizations of the electric field induced by the geomagnetic
  ($\hat{g}$) and Askaryan ($\hat{a}$) emission mechanisms are linear
  and follow the expected theoretical predictions (see
  Section~\ref{sec:groundpattern} and Fig.~\ref{fig:geoscheme-ask-geo-B-axis}):
\begin{equation}
\hat{g}=\frac{-\vec{V}\times \vec{B}}{|-\vec{V}\times \vec{B}|} ~~~ {\rm and} ~~~
 \hat{a}= \hat r
\label{eq:pols}
\end{equation}
where $\hat r$ is a radial unit vector that points from the antenna position to
the shower axis and is perpendicular to it, i.e. $\hat r$ is on the transverse
shower plane 
(perpendicular to the shower direction) that contains the antenna position. 

\item The amplitude of the electric field induced by each emission
mechanism separately is circularly symmetric in the transverse plane of the
shower, i.e. it depends only on distance to the shower axis.
Equivalently, the amplitude of the field on the ground for each mechanism is constant along ellipses
similar to the elliptical Cherenkov ring (see Section~\ref{sec:ring} and Fig.~\ref{fig:ellipse}).

We assume a radial symmetry for the amplitude of the electric fields in the
shower plane. However, when these fields are projected onto the ground, there are deviations 
from the assumed elliptical symmetry for the amplitudes. These deviations arise from differences 
in the shower-observer distances 
between the regions on the ground where the emission arrives early and late.
As shown later 
in Sections~\ref{sec:comparison} and \ref{sec:errors} they induce small inaccuracies in the prediction 
of the electric fields, and are not included in the model.

This symmetry property appears at the level of the ``raw" unfiltered traces of the pulses as shown in 
Fig.~\ref{fig:raw_traces}, where we plot the EW component (along $\vec V\times \vec B$) of 
$\vec{E}_{\rm geo}(t)$ and $\vec{E}_{\rm Ask}(t)$ in two antennas one located East and another West 
of the shower core on a circle in the shower plane. The geomagnetic traces 
are essentially on top of each other, while in the case of $\vec{E}_{\rm Ask}(t)$ 
they are also very similar but with opposite signs since the Askaryan EW component 
points in opposite directions East and West of the shower core. 

\begin{figure}
\begin{center}
\scalebox{0.35}{
\includegraphics{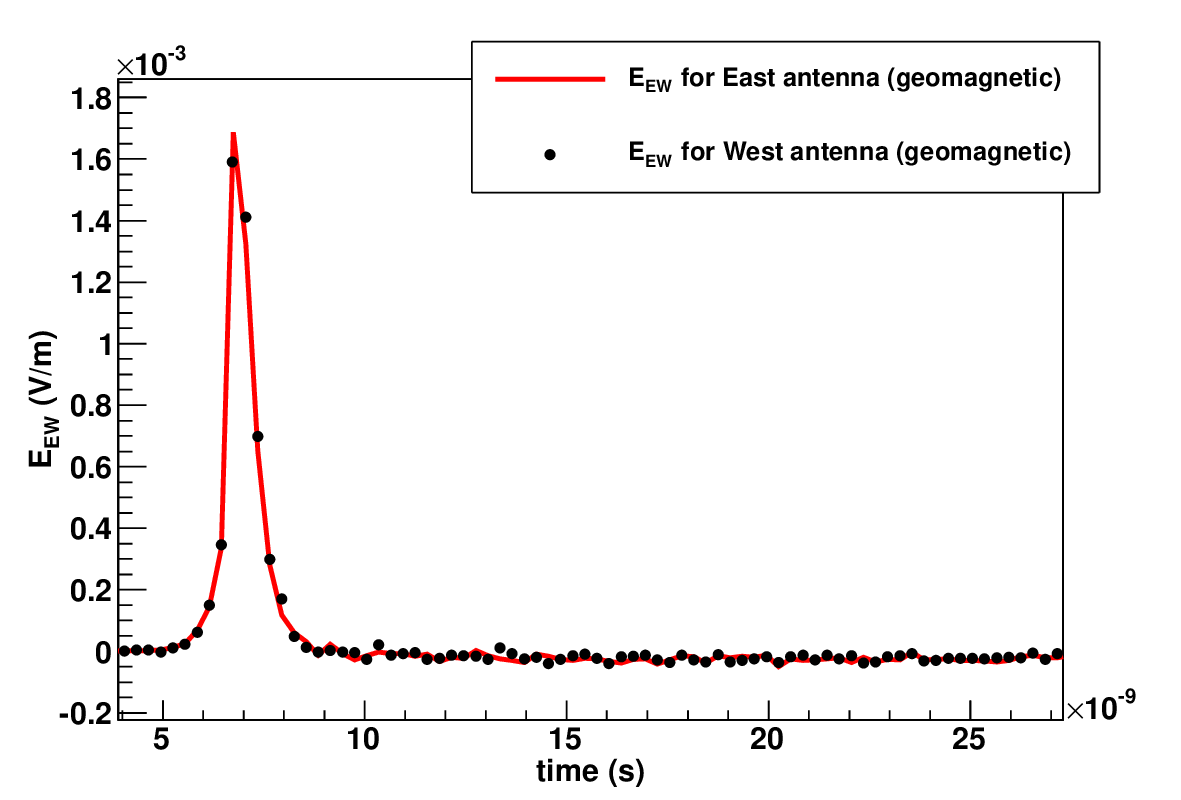}
\includegraphics{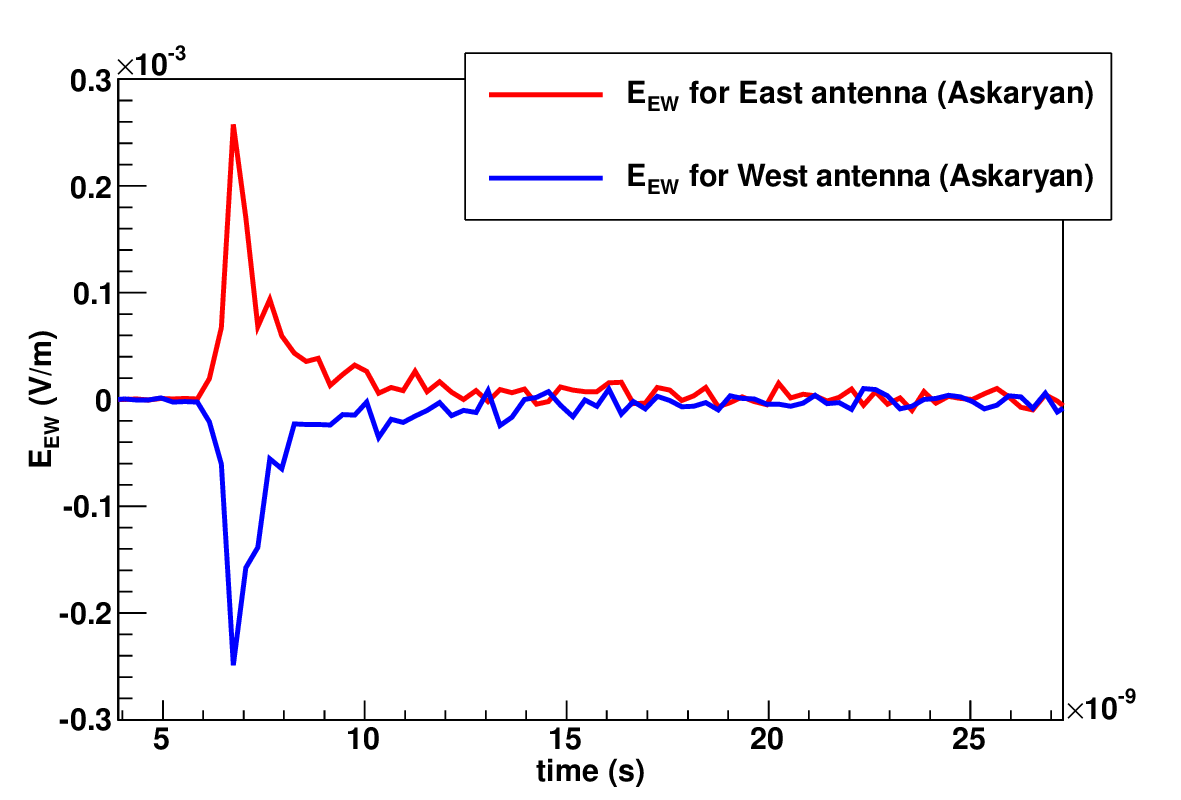}
}
\caption{
The EW component (i.e. along $\vec V\times \vec B$) of $\vec{E}_{\rm geo}(t)$ (left panel) and $\vec{E}_{\rm Ask}(t)$ (right panel)  
in two antennas located one East and another West of the shower core, on a circle in the shower plane.
The shower was simulated with ZHAireS and has $E=10^{17}$ eV, $\theta=30^\circ$ shower, arriving from North at the LOFAR site.
}
\label{fig:raw_traces}
\end{center}
\end{figure}

\item Both $\vec{E}_{\rm geo}(t)$ and $\vec{E}_{\rm Ask}(t)$ are accessible 
through Monte Carlo simulations of air shower development and associated radio emission 
such as those performed with the ZHAireS code described in \cite{ZHAireS}.

\begin{enumerate}[(C.1)]

\item 
$\vert\vec{E}_{\rm Ask}\vert(t)$ at any observer position, can be obtained in a shower simulation
where the magnetic field of the Earth is artificially switched off. In that case the 
Askaryan mechanism is the only one contributing significantly to the total electric field.
The assumption is:
\begin{equation}
\vec{E}_{\rm Ask}(t) = \vec{E}_{B_{\rm off}}(t)
\end{equation}

\item  The electric field induced by the geomagnetic mechanism 
can be approximately obtained by subtracting the electric field 
as given in a simulation performed with the geomagnetic field switched on
$\vec{E}_{B_{\rm on}}(t)$, 
from that obtained in the simulation with the magnetic field switched off 
described in the previous item.
For a particular observer position at time $t$:
\begin{equation}
\vec{E}_{\rm geo}(t)=\vec{E}_{B_{\rm on}}(t)-\vec{E}_{B_{\rm off}}(t)
\end{equation}
\end{enumerate}

This procedure is in essence approximate since it is not
possible to obtain exactly the same shower from simulations with and without the magnetic
field. By using the same random seed in both simulations we can only obtain similar showers.
To further investigate this, we have simulated two vertical showers, of the same energy,
arrival direction and random seed at the LOFAR site, but with the magnetic field switched on and artificially
switched off. For both showers we have calculated the average value $\langle y \rangle$ and RMS($y$) 
of the distribution of positions of electrons and positrons along the direction $\hat y$ parallel 
to the Lorentz force $\propto \vec V \times \vec B$ (with $\vec V$ parallel to the shower axis). 
In the shower with the magnetic field off $\langle y \rangle \sim 0$ m while the corresponding
value in the shower with the field on is slightly larger $\sim 0.5$ m but still close to 0. 
In the presence of the Lorentz force electrons deviate towards positive values of $y$ 
while positrons deviate in the opposite direction, inducing a negligibly small mean value of $y$ 
due to the excess of electrons over positrons in the shower (i.e. to the charge excess).
On the other hand, the RMS($y$) of the shower with magnetic field exceeds by $\sim 30$ m 
the corresponding RMS($y$) value of $\sim~120$ m for the shower without it.  
We have also checked that 
in the direction perpendicular to the Lorentz force, both the mean value of the perpendicular coordinate
and its RMS are equal in both showers within statistical uncertainties.  
At least for the geometry chosen and the magnetic field
at the LOFAR site, the presence of the magnetic field does not alter in a very significant manner 
the lateral structure of the shower. Also since the first interaction is exactly the same in 
both showers, the longitudinal development is similar.

\end{enumerate}


\subsection{Input and output observables. Methodology}
\label{sec:input}

\begin{figure}
\begin{center}
\scalebox{0.6}{\includegraphics{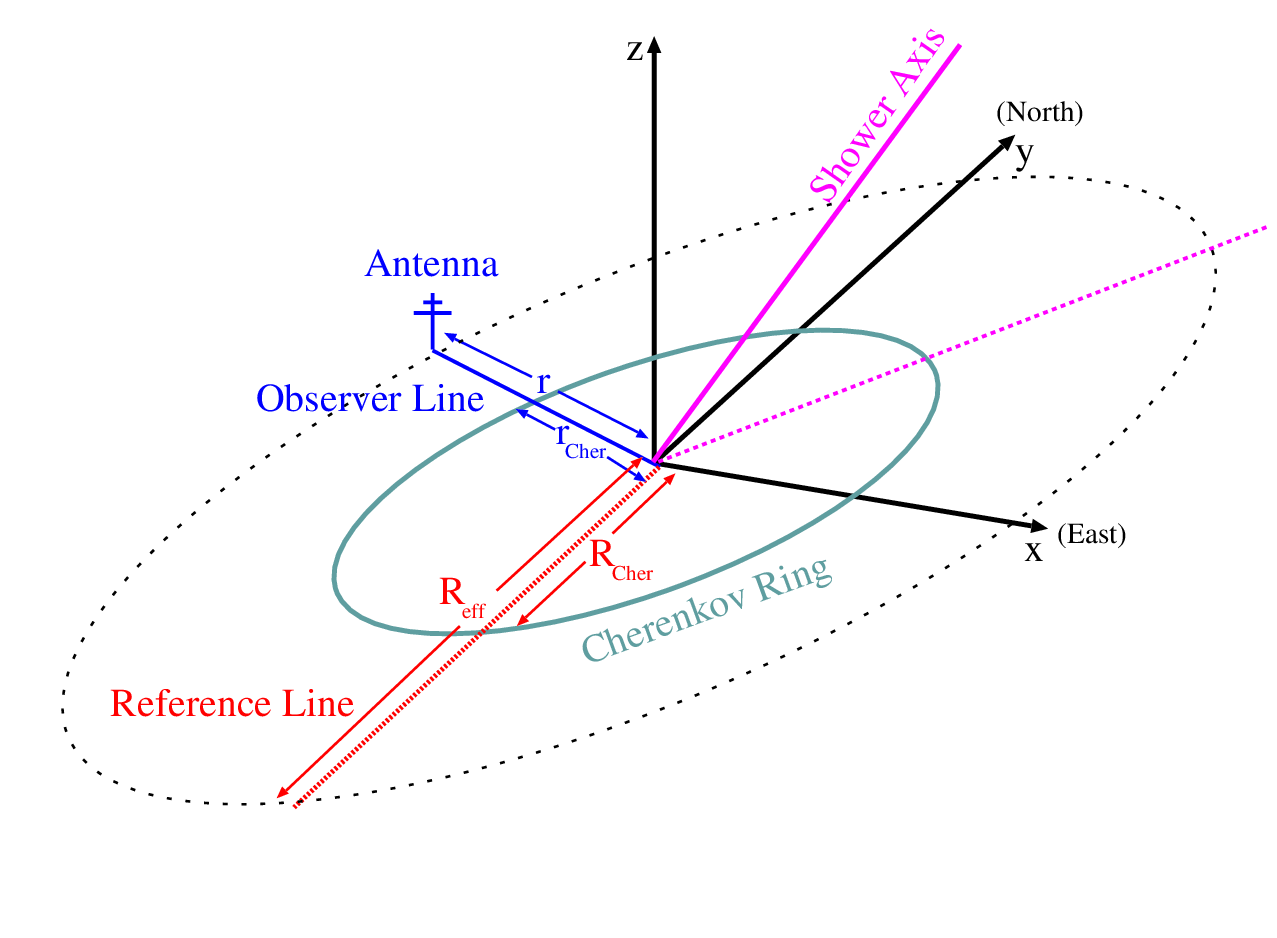}}
\caption{Sketch of the geometry of the shower-observer system used in this work. 
We show the shower axis hitting ground (xy plane), the Cherenkov ellipse on ground where the peak electric field
tends to be maximum and two lines: the ``reference line'' (in red, with distances to the core denoted 
as $R$), where the simulated electric field is used as input, and the
``observer line'' (in blue, with distances to the core denoted as $r$) where
we apply the model to predict the electric field.
We also show the distance to the Cherenkov ellipse along the reference line (denoted as $R_{\rm Cher}$) and
along the  observer line where the field is to be obtained (denoted as $r_{\rm
  Cher}$).}
\label{fig:ellipse}
\end{center}
\end{figure}

Our simple approach has been designed to predict a time independent 
vector observable $\vec {\mathcal{E}}$ related to the electric field and
parallel to it. 
The model was not developed with the aim of predicting the raw and/or filtered 
electric field trace vs time, although in principle it could be extended to 
reproduce them. This does not represent a limitation of our 
approach since many experiments actually use time independent measurements
of the electric field (see for example \cite{AERA}). 
The only inputs in our approach are the amplitudes of the observable for the geomagnetic 
and Askaryan emission mechanisms separately at a few observer positions on ground. 
These amplitudes will be denoted as ${\mathcal E}_ G$ and ${\mathcal E}_ A$.
The polarization patterns are part of the model. These positions are chosen at
different distances $R_{i}$ with respect to the shower core along a line, 
which hereafter will be referred to as the {\it reference line}.
The input amplitudes for antennas along the chosen reference line
are extracted from the full Monte Carlo simulations (see assumption C in
Section~\ref{sec:assumptions}). As will be shown later in this work, the
actual direction chosen for the reference line is practically irrelevant, and we 
obtain accurate predictions for any reference line used as input. A sketch of the 
shower-observer geometry is shown in Fig.~\ref{fig:ellipse} where an 
example reference line is also depicted.  

It is straightforward to extrapolate the amplitudes ${\mathcal E}_G(R_i)$ and ${\mathcal E}_A(R_i)$  
to any observer position ($r,\phi$) and to obtain
their expected polarizations following the assumptions described in
3.1.A. By adding these two contributions we can obtain
the net electric field ${\vec E}(r,\phi)$, where $r$ and $\phi$ are
cylindrical coordinates in the ground plane. 

Firstly, from the actual depth of maximum shower development $X_{\rm  max}$ 
obtained in the simulation, we define the elliptical
Cherenkov ring on the ground as given in Eq.~(\ref{eq:calcring}).
To obtain the amplitude of a given contribution at a given antenna
position $(r,\phi)$ we draw a line joining it to the impact point of the
shower axis, which we can refer to as the ``observer line''. 
The amplitudes are assumed 
constant along ellipses similar to the Cherenkov ring (see Fig.~\ref{fig:ellipse}), i.e.:
\begin{equation}
{\mathcal E}_ G(r)={\mathcal E}_ G(R_{\rm eff}),~~~~{\mathcal E}_ A(r)={\mathcal E}_ A(R_{\rm eff}),
\end{equation}
where $R_{\rm eff}$ is simply obtained from the following geometrical relation arising from 
the similarity of the ellipses depicted in Fig.~\ref{fig:ellipse}\footnote{The choice of the positions of 
the elliptical Cherenkov ring along the reference 
and observer lines, although natural, is  
arbitrary due to the assumed elliptical symmetry of the amplitudes on the ground.}:
\begin{equation}
\frac{R_{\rm eff}(r,\phi)}{R_{\rm Cher}}~=~\frac{r}{r_{\rm Cher}(r,\phi)}
\label{eq:similarity}
\end{equation}

Secondly, the expected polarization 
of the geomagnetic (${\hat{g}}$) and Askaryan (${\hat{a}}$) fields  which depend on the
direction of the shower axis ($\theta,~\varphi$) and on 
the position of the observer, are obtained 
with the aid of Eq.~(\ref{eq:pols}). The resulting field is then given by\footnote{We use a simple linear
  interpolation to obtain the amplitudes at any given $R_{\rm eff}$ from
  the few antenna positions $R_i$ used as input.}:
\begin{equation}
\vec{E}(r,\phi) = {\mathcal E}_ G(R_{\rm eff})~{\hat{g}} + {\mathcal E}_ A(R_{\rm eff})~{\hat{a}}
\label{eq:field}
\end{equation}
This expression applies regardless of the choice of observable,
the only restriction is that it should be a time independent measure of the electric field.
In fact, several observables can be constructed from the full bandwidth pulse
$\vec{E}(t)$ and different observables are used in each experiment to describe
the electric field at ground level. Typically these observable definitions
take into account detector characteristics such as its working frequency
band. Example choices are the average or the maximum of the electric field amplitude of the bandpass filtered pulse.

For concreteness we choose an observable throughout this work similar to that used
in the AERA experiment \cite{AERA}, described in the
following. For each antenna position $R_i$ along the reference line, we first
apply a band pass filter between 30 and 80 MHz to the simulated full
bandwidth pulses $\vec{E}_{\rm geo}(R_i,t)$ and $\vec{E}_{\rm Ask}(R,t)$\footnote{
This bandwidth is currently being exploited in air shower radio detection experiments
such as AERA \cite{AERA}, LOFAR \cite{LOFAR} and CODALEMA \cite{CODALEMA}.}. 
We use the amplitude of the analytical signal, which is obtained by a Hilbert transform
of the filtered pulses and calculate
the time dependent amplitudes for the geomagnetic,  
$\vert\vec{\mathcal H}_G\vert(R_i,t)$, and the Askaryan,
$\vert\vec{\mathcal H}_A\vert(R_i,t)$, components. To obtain time independent
quantities we finally take the maxima of these amplitudes at the
selected points along the reference line, 
${\mathcal E}_G(R_i)={\rm max}(\vert\vec{\mathcal H}_G\vert(R_i,t))$ and
${\mathcal E}_A(R_i)={\rm max}(\vert\vec{\mathcal H}_A\vert(R_i,t))$.

\section{Results}
\label{sec:results}

Here we compare the predictions obtained using the methods
described in this work with the results of full simulations of the radio
emission performed with the ZHAireS code \cite{ZHAireS}.
We demonstrate the validity of the assumptions of our
model, exploring its range of applicability.

\subsection{Monte Carlo simulations with ZHAireS}
\label{sec:sims}

Using the ZHAireS code \cite{ZHAireS}, we have produced sets of shower
simulations with the parameters of the Pierre Auger Observatory site (LOFAR site)
namely: ground altitude 1400 m (10 m) above sea level, and geomagnetic field of intensity 
$23 \; \mu {\rm T}$ ($49.25 \; \mu {\rm T}$) and inclination $-37^\circ$
($67.8^\circ$).    

The sets of simulations contain proton-induced showers at $E=10^{17}$ eV arriving   
from the North 
$\varphi=90^\circ$ for both Auger and LOFAR sites, and from 
the West $\varphi=180^\circ$ (the South $\varphi=270^\circ$) for the Auger (LOFAR) site. 
For each azimuthal angle we have simulated 
showers with several zenith angles ranging from $\theta=0^\circ$ to 
$\theta=80^\circ$ in steps of $\Delta\theta=5^\circ$.
We placed several observers at different distances from the shower core along 8 directions
on ground, namely towards magnetic North (N), South (S), 
East (E), West (W), NorthEast (NE), NorthWest (NW), SouthEast (SE) and SouthWest (SW). 
Each shower geometry was simulated twice using the same random seed: once
with the magnetic field switched on and once with it artificially switched off.
The electric field induced by the geomagnetic and Askaryan 
mechanisms was obtained (as explained in Section \ref{sec:model}) for each observer position
and for each shower geometry. 

A subset of the simulated pulses  
i.e. those at the positions that lie along a reference line, were used to
construct the input of the model for each shower geometry
 (see Section \ref{sec:model}). These are used to predict the
electric field at any other position away from the reference line, and in
particular at the same positions where the pulses were simulated so that the comparison between the net field inferred from our method and
that from the full simulation could be done. 
We calculated the same observable described in section \ref{sec:input} from the 
pulses obtained from full simulations.

\begin{figure}
\begin{center}
\scalebox{0.65}{\includegraphics{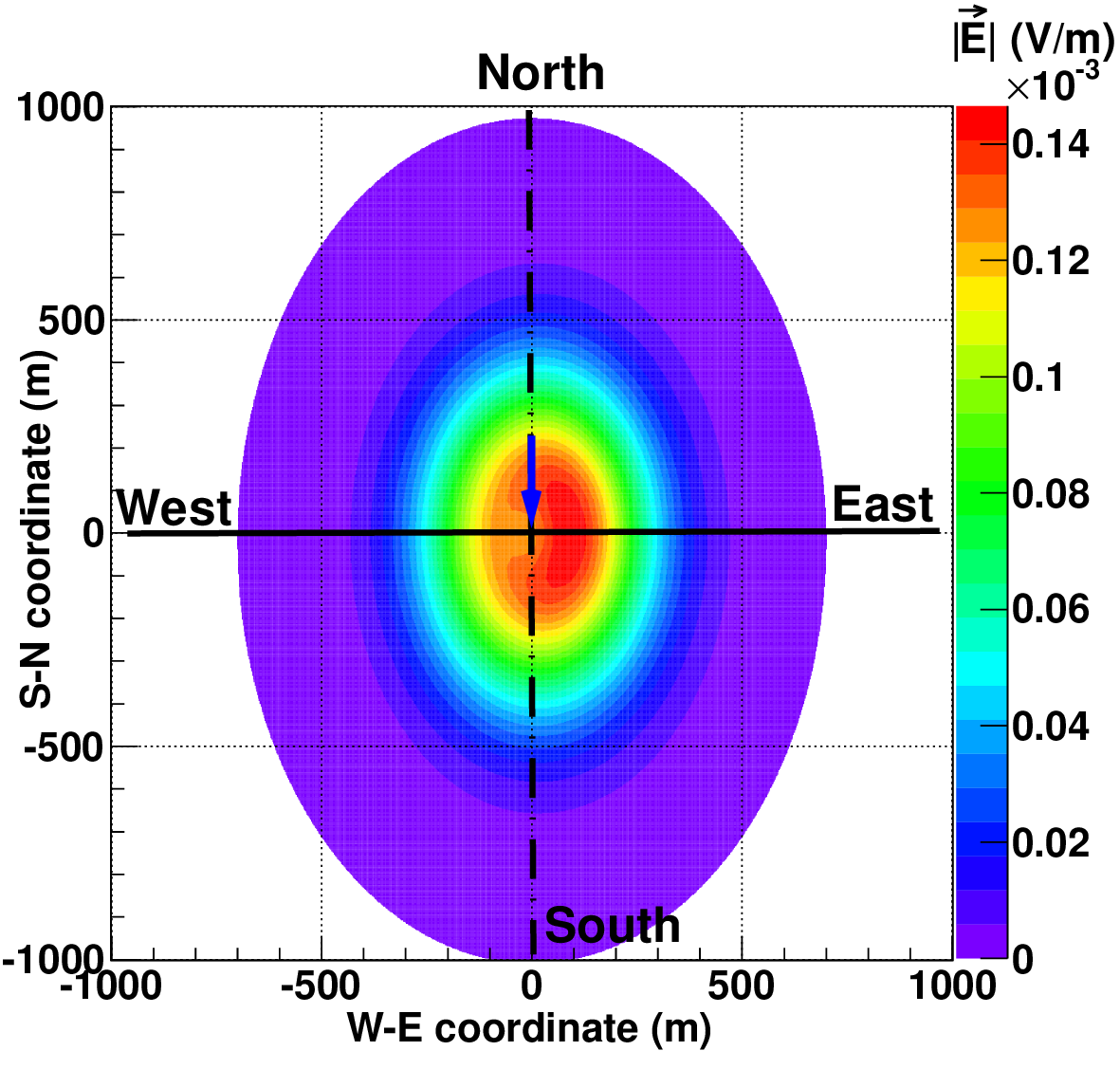}}
\caption{Two-dimensional pattern of the amplitude of the electric field on the
  ground predicted by the simple model for a $10^{17}$ eV, $\theta=45^\circ$ 
  shower arriving from magnetic North at the LOFAR site. We used 36 antenna 
  positions as input to our approach, which are then used to estimate the signal 
  in antennas at every 2 m in a rectangular grid (a total of $\sim 2\times10^5$ antennas).}
\label{fig:map}
\end{center}
\end{figure}

As a first example we show in Fig.~\ref{fig:map} the two-dimensional pattern 
of the electric field on the ground for a $\theta=45^\circ$ shower arriving from magnetic North
at the LOFAR site. The amplitudes of the electric field at the antenna positions East 
of the core are used as input, and the electric field is obtained with the model
at all other positions ($\sim 2\times10^5$ in total). The East-West asymmetry as well
as the Cherenkov ring can be seen.

\subsection{Predictions compared to full simulation}
\label{sec:comparison}

With the aid of the simulations described before we tested the validity
of our method to predict the total electric field at 
positions away from the reference line used as input.

\begin{figure}
\begin{center}
\scalebox{0.7}{\includegraphics{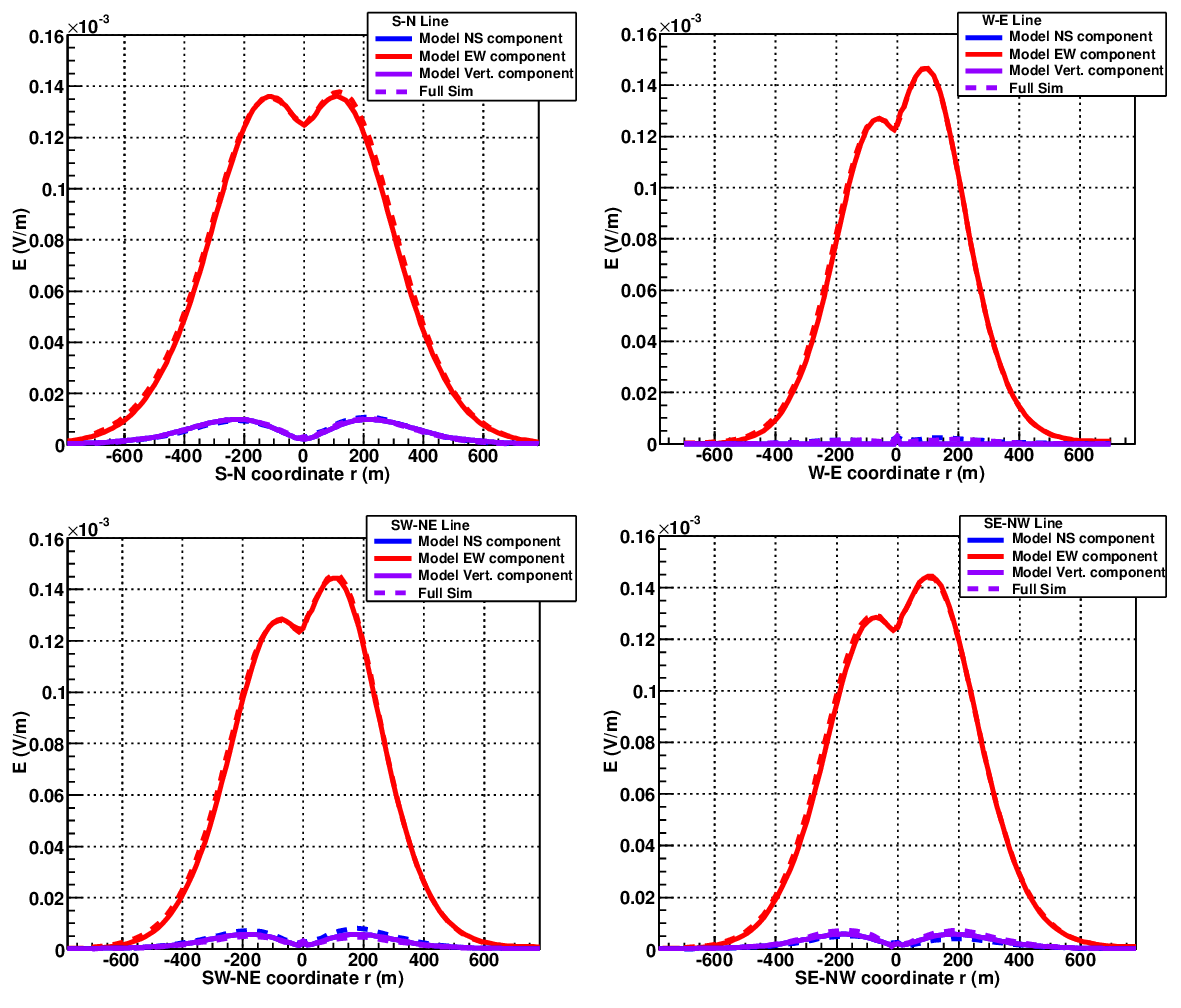}}
\caption{Comparison of the components of the electric field as predicted by
  the model (solid lines) with those obtained in ZHAireS simulations (dashed lines) 
as a function of distance to the core $r$ for observer positions along several directions
on ground: South to North (top left), West to East (top right), SouthWest to NorthEast 
(bottom left) and SouthEast to NorthWest (bottom right). 
The geometry corresponds to a proton-induced shower of energy $E=10^{17}$ eV 
coming from the North at $\theta=45^\circ$ at the LOFAR site.
We used as input the amplitude of the electric field from the simulations at positions along the 
East direction (the reference line) and predicted the field components
along all other lines shown in the plot. Negative values of $r$ refer to S, W, SE and SW directions.}
\label{fig:fullcomparison}
\end{center}
\end{figure}

We found a very good agreement between the predictions of the simple model 
and the Monte Carlo simulations in the vast majority of shower geometries. 
As an example, using the same configuration as in Fig.~\ref{fig:map},
we compare in Fig.~\ref{fig:fullcomparison} the components of electric field 
as obtained in the model with those in full simulations. 
In this particular example we used the line of antennas
East of the core as the input reference line. 
The agreement is within a few percent for
all distances to the shower core $r$ and all directions. The shower geometry used in
Fig.~\ref{fig:fullcomparison} results in the angle $\alpha$ between $\vec{V}$
and $\vec{B}$ of $\sim112.8^\circ$  or equivalently $\sin\alpha\sim 0.92$. 
As a consequence the geomagnetic contribution dominates and  
the total electric field is dominated by the EW component (red
lines), which is the direction of $-\vec{V} \times \vec{B}$  in this geometry.

The positions where the field reaches a maximum follow
the expected elliptical shape of the Cherenkov ring. 
As expected for a shower coming from North, the
major axis of the elliptical Cherenkov ring lies along the NS direction, while
the minor axis lies along the EW direction. Also visible in
Fig.~\ref{fig:fullcomparison} is the expected EW asymmetry in
the value of the EW component of the field, which arises from the interference
of the geomagnetic and Askaryan contributions. In the case
of antennas that lie in the W-E line (top right panel of
Fig.~\ref{fig:fullcomparison}), the geomagnetic and Askaryan polarizations are
both approximately parallel to the W-E line and point in the same direction for
observers East of the shower core, and in the opposite direction for observers
West of the core, leading to the asymmetry. 

\begin{figure}
\begin{center}
\scalebox{0.65}{\includegraphics{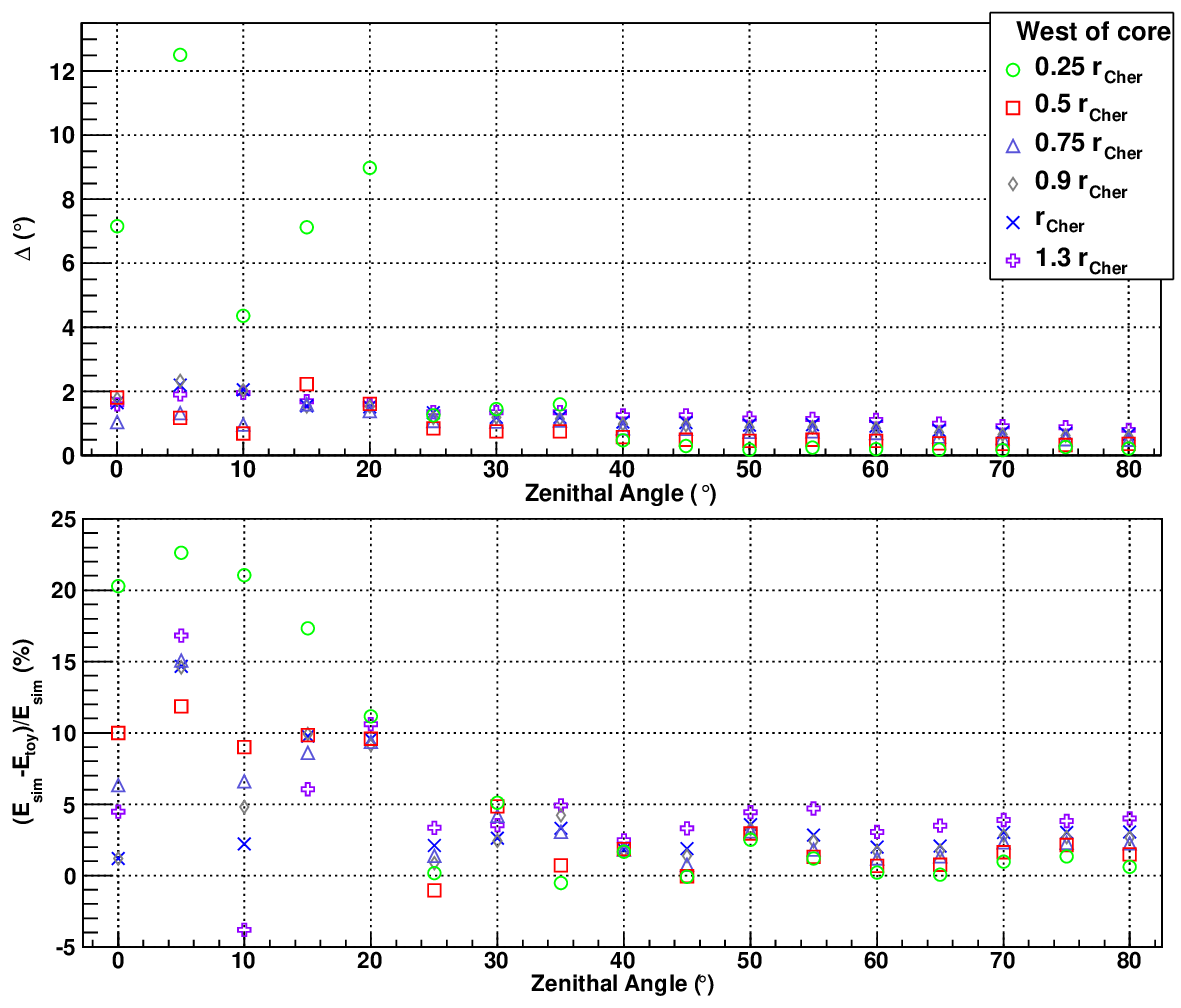}}
\caption{Top: opening angle $\Delta$ between the total electric field
 predicted by the simple model and that obtained 
in the full simulation. The opening angle is shown as a function of 
shower zenith angle $\theta$ for several distances from the core and for observer positions along
the West direction on the ground.
The distances to the shower core are scaled to the distance to the
Cherenkov ring $r_{\rm Cher}$ at each $\theta$.
The results correspond to proton-induced showers of energy $E=10^{17}$ eV coming from the North at the LOFAR site. 
Observer positions along the East direction, not shown in the plot,
were used as input.
Bottom: relative difference between 
the amplitude of the net electric field as obtained with our method and that from the full 
simulation for the same observers and shower geometries as in the top panel.
}
\label{fig:AngleDeltaE}
\end{center}
\end{figure}

We have extensively tested our method for all the simulated showers. Some of these comparisons are summarized in
Fig.~\ref{fig:AngleDeltaE}, where we show the opening angle
between the polarizations of the 
electric field predicted by the model and that obtained in the 
full simulation. The opening angle is shown as a function of 
shower zenith angle $\theta$ for observer positions along
the West direction on the ground at several distances to the shower
core. For each $\theta$ these distances are given relative to the
distance $r_{\rm Cher}$ to the Cherenkov ring, e.g. $0.5\,r_{\rm Cher}$ (red squares)
refer to positions half way between the core and the Cherenkov ring. 
It is important to notice that $r_{\rm Cher}$ increases with $\theta$,
and in this plot equal values
of $r/r_{\rm Cher}$ for different $\theta$ correspond to different
actual distances on the ground $r$.
As can be seen in the figure, for showers 
with  $\theta\gtrsim20^\circ$ the
opening angle is very small, below $\sim 2^\circ$. In the bottom panel of Fig.~\ref{fig:AngleDeltaE} we show the relative difference between 
the amplitude of the total electric field 
as obtained with our method and that from the full 
simulation for the same observers and shower geometries as in the top panel.
The differences are smaller than $\sim 5\%$ for $\theta\gtrsim20^\circ$. These
comparisons demonstrate that the method works to a very good level of
accuracy, validating the assumptions we made in Section \ref{sec:model}. The
more vertical showers with $\theta\lesssim 20^\circ$ will be discussed in
Section~\ref{sec:errors}. 

\begin{figure}
\begin{center}
\scalebox{0.65}{
\includegraphics{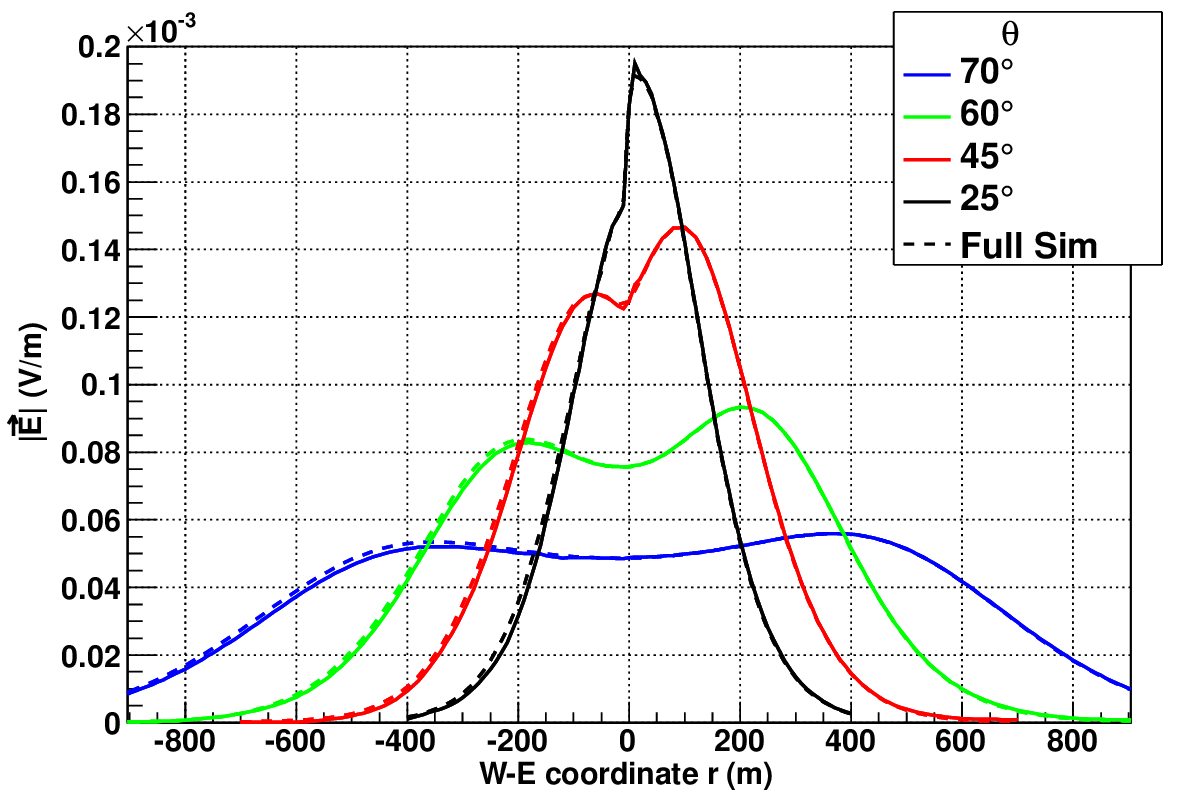}
}
\scalebox{0.65}{
\includegraphics{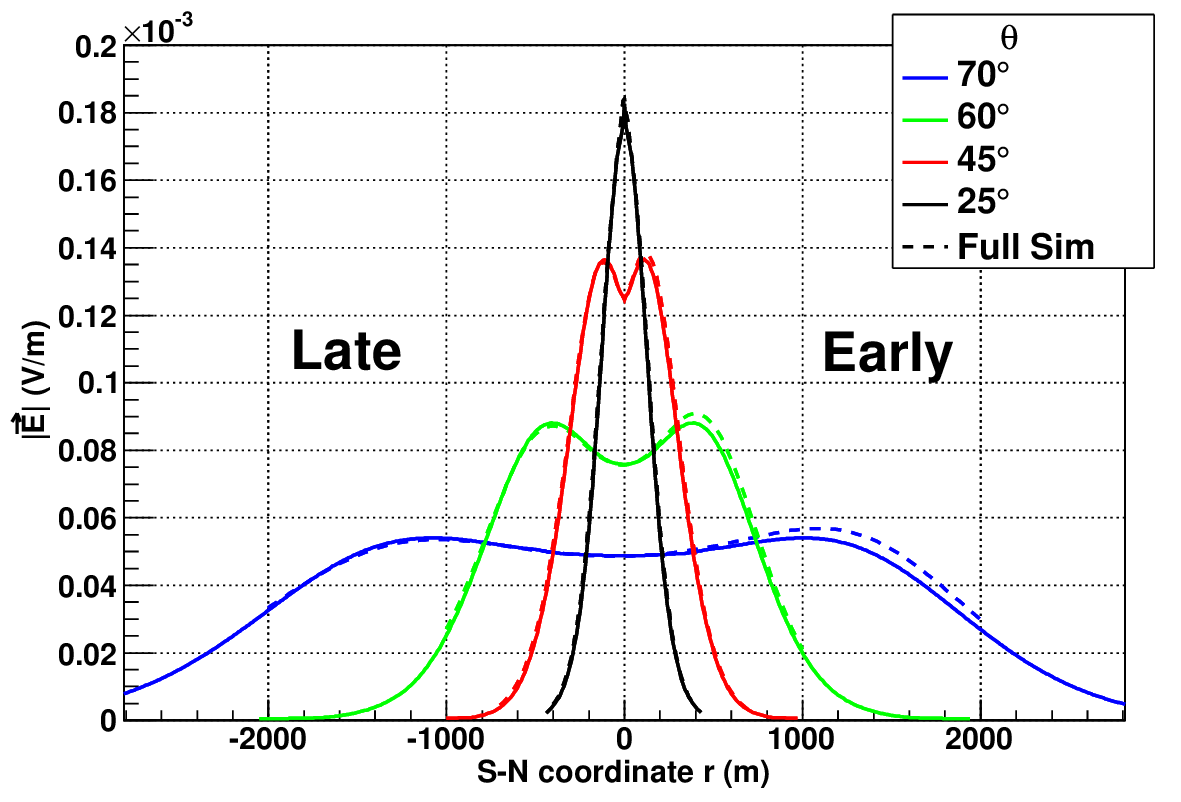}
}
\caption{
Comparison of the modulus of the electric field $\vert \vec{E}\vert$ as
predicted by the simple model (solid lines) with that obtained in ZHAireS simulations (dashed lines) 
as a function of distance to the core $r$ for observer positions along
the West to East (top) and South to North (bottom) directions on ground.   
The geometry corresponds to proton-induced showers of energy $E=10^{17}$ eV 
coming from the North at different $\theta$ (see legend) at the LOFAR site.
We used the amplitude of the electric field from the simulations on the positions along the 
East direction as inputs (the reference line) and predicted the field components
along all other lines shown in the plot. Negative values of $r$ refer to W (in the top panel) 
and S (bottom) directions.}
\label{fig:cathedrals}
\end{center}
\end{figure}

In Fig.~\ref{fig:cathedrals} we show a comparison between $\vert \vec{E}\vert$ 
as predicted by the model and as obtained in Monte Carlo 
simulations for showers at different zenith angles and along two directions
on the ground, namely the EW (top) and NS (bottom) directions. Again,
the agreement between our method and the simulations is at a few percent
level. The Cherenkov ellipse is apparent, 
the major axis of the ellipse lying along the NS direction as expected for a shower coming
from the North. The major and minor axes of the ellipse grow with increasing $\theta$,  
mainly due to geometrical projection effects associated to the fact that the shower develops further 
from ground as $\theta$ increases. This is only partially compensated by a decrease of the Cherenkov angle because  
as the zenith angle increases, shower maximum occurs in a less dense atmosphere \cite{ZHAireS_ANITA}. 
In the most vertical shower shown in Fig.~\ref{fig:cathedrals} the Cherenkov ellipse is less apparent, 
mainly because in less inclined showers the lateral extent of the shower
becomes important \cite{Washington_icrc2013}.
As explained above, the asymmetry of the field along the EW direction is
clearly visible in the top panel of Fig.~\ref{fig:cathedrals}. 
In the bottom panel of
Fig.~\ref{fig:cathedrals} one can see that the amplitude of the electric field
North of the core is slightly underestimated. In this geometry
($\varphi=90^\circ$), antennas North of the core are closer to the shower
than antennas South of the core, leading to larger fields North of the
core. 

\begin{figure}
\begin{center}
\scalebox{0.65}{\includegraphics{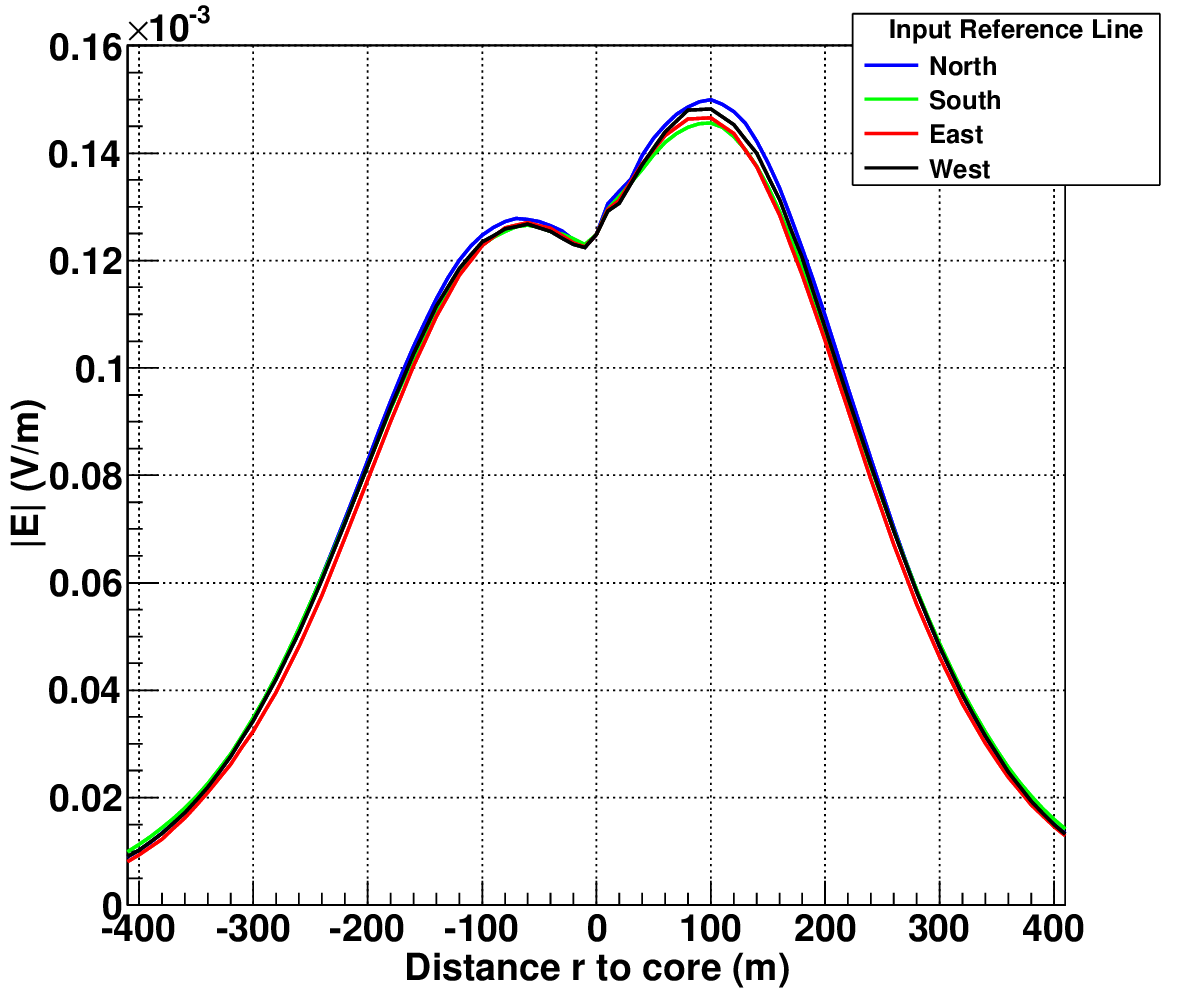}}
\caption{
Modulus of the total electric field $\vert \vec{E}\vert$ as a function of distance to core
along the W-E line (for the same geometry as in Fig.~\ref{fig:fullcomparison}). 
The electric field was obtained from the simple model  
using as input the electric field simulated along 4 different
reference lines one at a time, namely using observer positions
along the North, South, East and West directions from the shower core.}
\label{fig:RefLinesymmetry}
\end{center}
\end{figure}

As stated earlier in this work, the choice of  
the reference line, where the geomagnetic 
and Askaryan electric fields are simulated and used as input to the model, is mostly irrelevant. 
This is illustrated in Fig.~\ref{fig:RefLinesymmetry},
where we show $\vert \vec{E}\vert$ as a function of distance to core
along the W-E line for the same shower geometry as in Fig.~\ref{fig:fullcomparison}. 
Each line in Fig.~\ref{fig:RefLinesymmetry} represents the electric
field obtained from the simple model using as input one of 4 different reference
lines (see legend).
One can see that there is
only a small dependence on the choice of the reference line, which is mostly
visible in the right hand peak around $r\simeq100$ m. 
The largest differences between the fields are obtained when 
comparing the N and S input reference lines. They can be  
understood in terms of early/late shower effects. 
Since the shower comes from the North, observers North of
the core are closer to the shower than those at the South, 
inducing larger signals in the Northern reference line than in the Southern one.
This effect is propagated to the predicted field amplitude of the model, 
which is $3\%$ larger when the North direction is used as 
input compared to that obtained with the South line as reference, for this
particular shower geometry.  When the
West or East directions are used as input, the fields
tend to fall in between the predictions obtained using the North and South 
reference lines as could be expected, which is particularly clear at
the Cherenkov peaks. This early/late effect diminishes near the
core, where the N and S antennas have equivalent distances to 
points along shower axis. On the other hand the effect increases as the shower zenith angle $\theta$
rises, since differences in distance to shower maximum from observers
located N and S of the core become much more relevant.

\subsection{Sources and effects of inaccuracies}
\label{sec:errors}

Our simple model predicts the electric field to a few percent when compared
to full simulations in the vast majority of shower-observer geometries
as shown in Section~\ref{sec:comparison}. 
However, in a few particular cases the inaccuracies in the predicted field become more apparent. 
The main sources of inaccuracies arise from deviations from the
assumptions listed in Section~\ref{sec:model} and the extent to which each of these deviations affects
the predicted electric field depends on the
geometry of the shower/observer system.

{One of them is related to the input of the model and the 
way in which separate the Askaryan and geomagnetic contributions to the
electric field amplitudes (assumption C.2. in Sec. 3.1.). 
These could  arise because the spatial structure of the shower changes as we turn off 
the magnetic field in the simulation and because of small differences in the time structure 
of the geomagnetic and Askaryan pulses. 
However, since in most cases we get a very good approximation of the full
simulation by using the separate contributions, these errors must be small.

Early/late effects 
induce only very small inaccuracies in the predicted 
electric field  in the case of low zenith angle showers, but become 
larger as the zenith angle increases ($\sim 5\%$ at the peak for the
$70^\circ$ shower shown in the bottom panel of Fig.~\ref{fig:cathedrals}).
 
We have checked that the geomagnetic polarization assumed in
Eq.~(\ref{eq:pols}) describes well that obtained in full Monte Carlo
simulations at all relevant distances to the shower core. However this is not
the case for the Askaryan polarization at small distances to the core and
manifests into larger opening angles between the expected and fully simulated
polarizations. This makes our model less accurate for the prediction of the
polarization of the net electric field in vertical showers $\theta\lesssim
20^\circ$ in the case of observers near the shower core at distances $r=0.25~r_{\rm
  Cher}\sim 20$ m as can be seen in Fig.~\ref{fig:AngleDeltaE}. 

Inaccuracies also arise in the electric field amplitude predicted by the model due to the vector addition of the Askaryan and geomagnetic
polarizations (Eq.~\ref{eq:field}). In fact the model is less accurate when the shower-observer geometry is such
that the geomagnetic component of the emission is 
either smaller than the Askaryan component or has approximately the same 
magnitude and opposite polarization.
For the geometries of the simulated showers used in
this work, this occurs along the W-E line of antennas, either West or East of
the core. In those geometries the model predicts 
a small total electric field due to the cancellation of the geomagnetic and
Askaryan contributions. Several examples are shown in the right panels of
Fig.~\ref{fig:badcases} and represent some of the worst
agreements between model and simulation of all studied showers.  For instance, in the particular case of a shower coming from the
North with $\theta=60^\circ$ and observers along the West to East line at the Auger site
depicted in the bottom right panel, the expected Askaryan
and geomagnetic polarizations are both horizontal (parallel to the W-E line),
and point in opposite directions East of the core. The geomagnetic contribution
dominates the emission up to $r/r_{\rm Cher}\simeq 1$, where it is equal to
the Askaryan one, while for larger distances the Askaryan mechanism is
slightly dominant. In this scenario, the electric field cancels in the model at a distance $r/r_{\rm Cher}\simeq 1$ (East of the core). 
In contrast, this
perfect cancellation does not occur in the simulation, as can be seen in the
same figure. However, although the differences between the model and full simulations are
quite evident, they are still small if compared to the peak value of $\vert\vec{E}\vert$ for this
shower (below $\sim 5\%$).  
These differences may be attributed to small deviations in the polarization 
in conjunction with the inherent and unknown uncertainties on the amplitude of the geomagnetic field 
due to assumption C.2 in Section 3.1.
Other examples are shown in the top and middle right panels, where the
Askaryan and geomagnetic fields point in opposite directions West of the core.
As can also be seen in Fig.~\ref{fig:badcases}, these differences disappear elsewhere on the ground, where
the Askaryan and geomagnetic polarizations do not have approximately opposite
directions, such as West (East) of the core in the bottom (top and middle) right panel 
and along the whole S-N line of antennas depicted in the left panels. This
means that these inaccuracies only affect observers on a very small area on the
ground.  In some cases the discrepancies between
the assumed and fully simulated polarizations are small, but there are also
regions near the core in which there is a large discrepancy in the Askaryan polarization. That can cause large errors in the electric
field amplitudes in these regions as shown in the bottom panel of
Fig.~\ref{fig:AngleDeltaE} for $\theta\lesssim 20^\circ$. 
In summary, these larger inaccuracies occur only if the
Askaryan contribution is similar to or dominates over the geomagnetic emission,
and only in regions on the ground where the two mechanisms have almost
opposite polarizations.

\begin{figure}
\begin{center}
\scalebox{0.35}{
\includegraphics{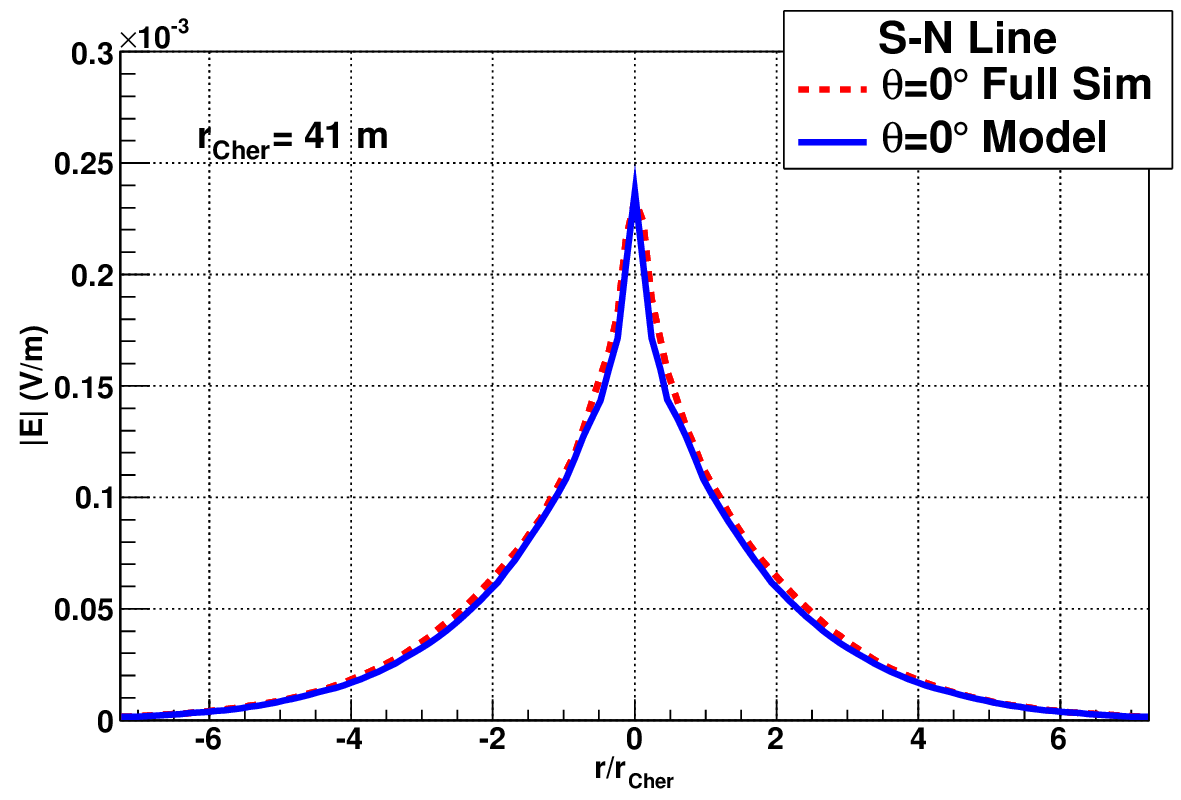}
\includegraphics{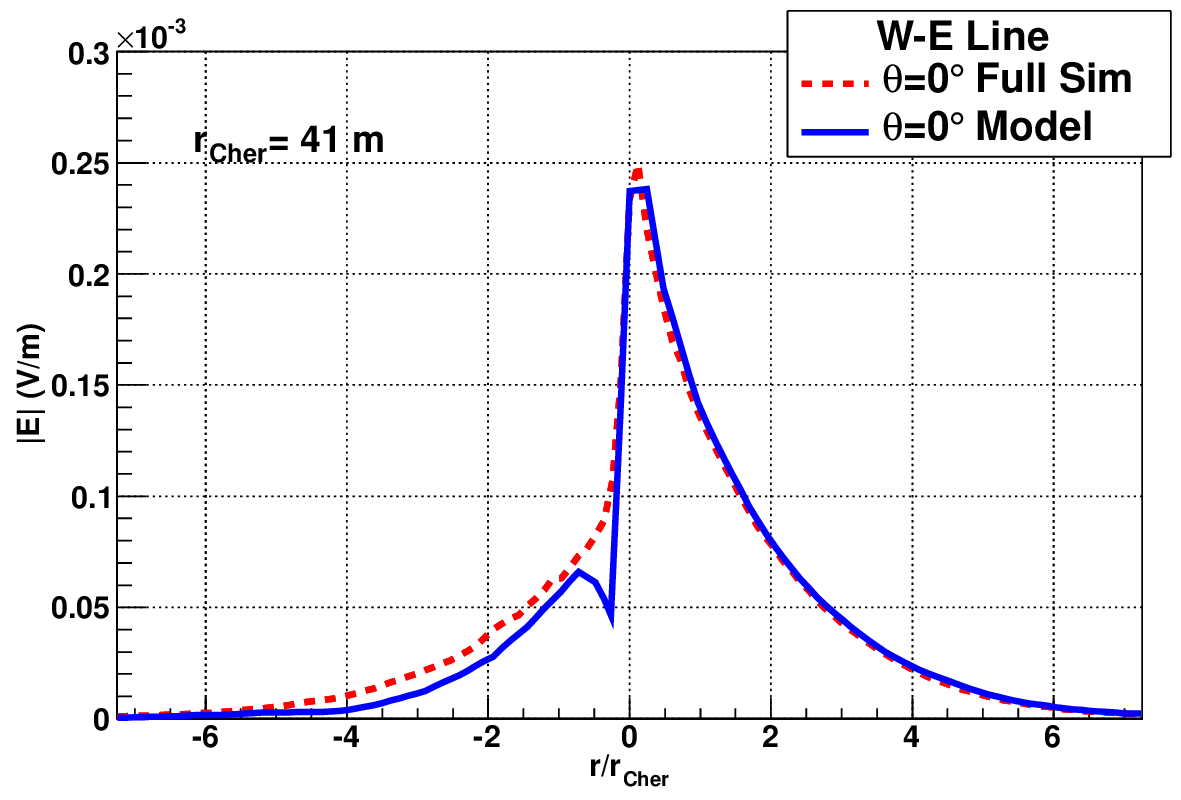}
}
\scalebox{0.35}{
\includegraphics{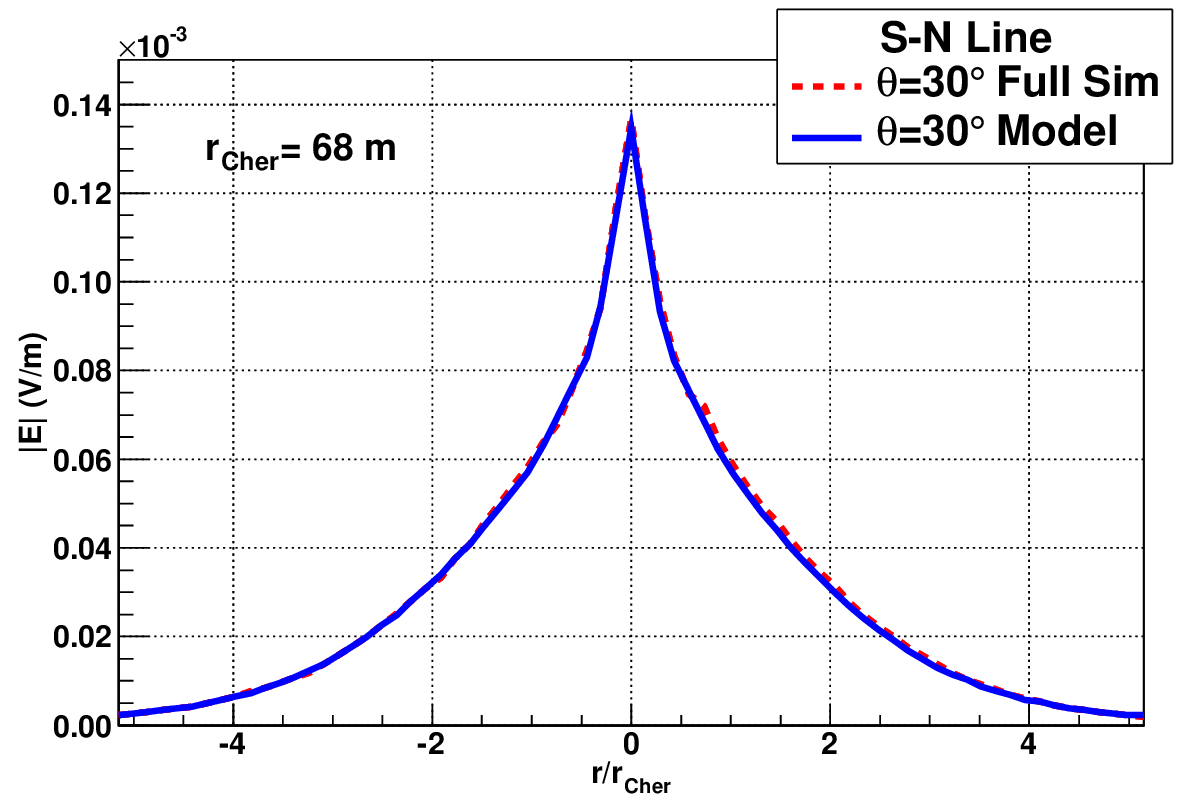}
\includegraphics{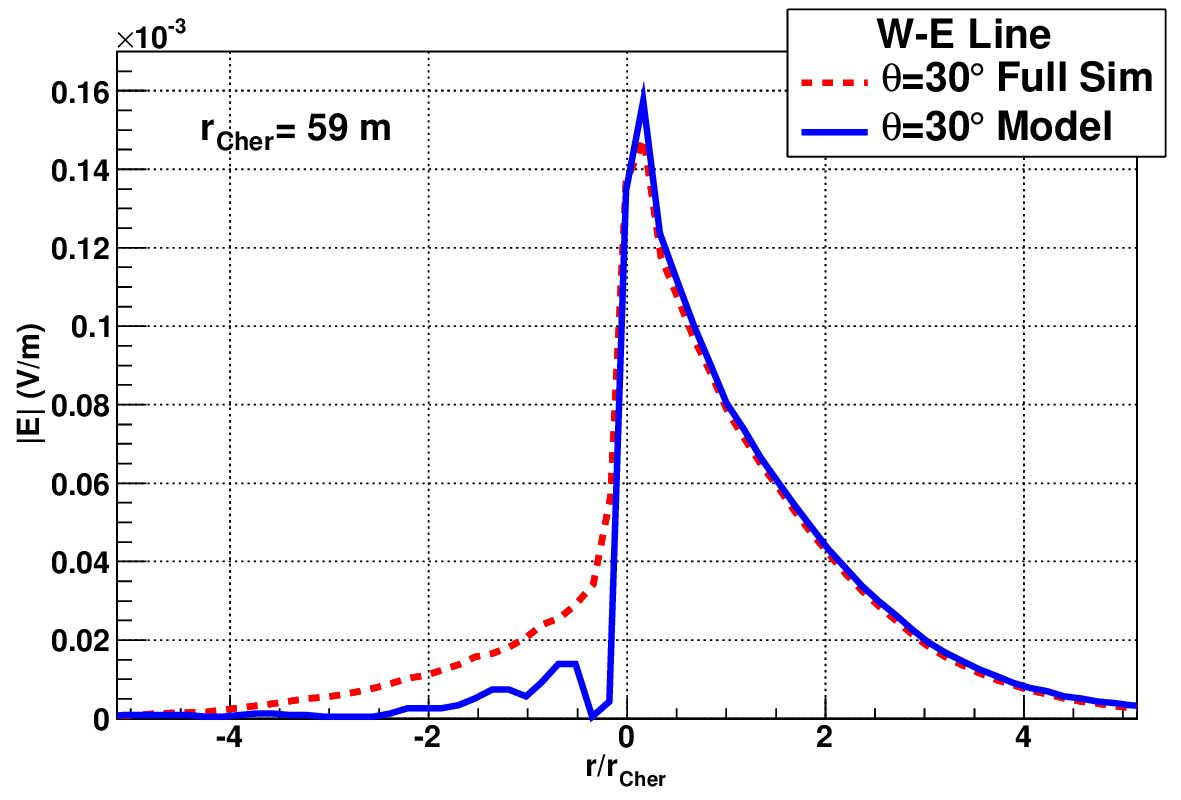}
}
\scalebox{0.35}{
\includegraphics{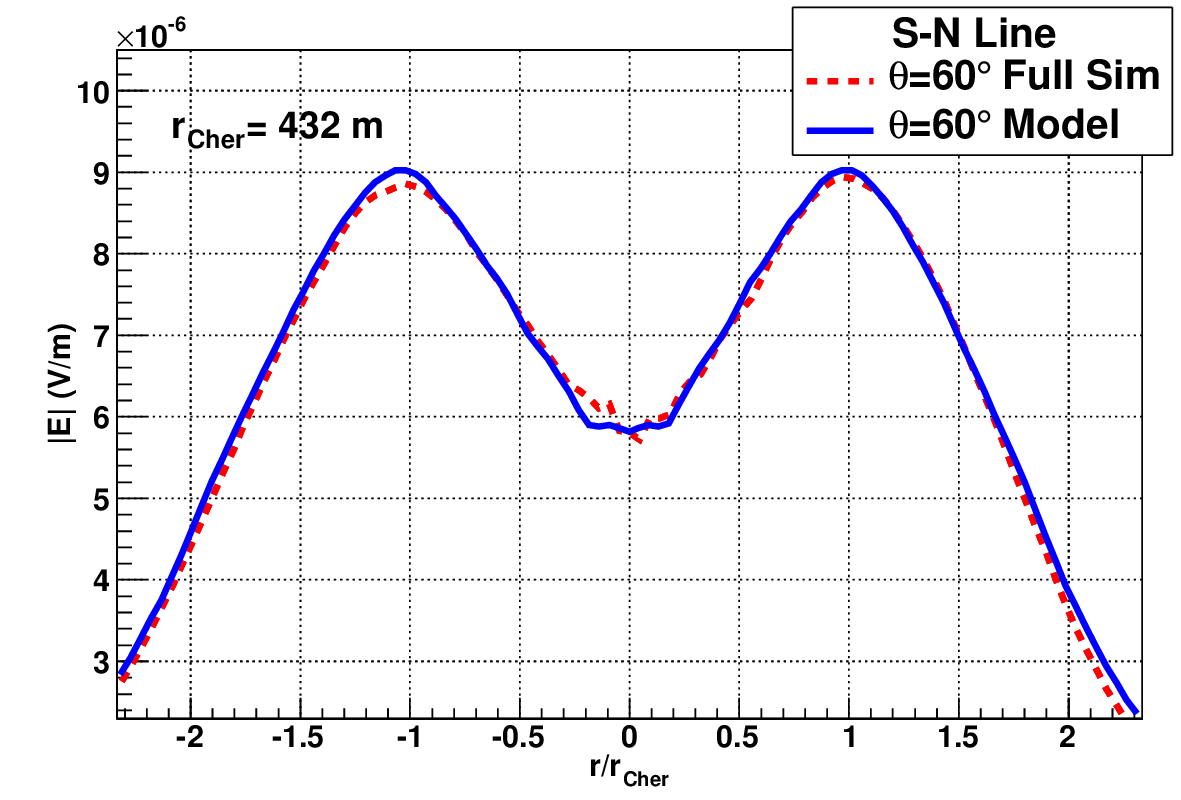}
\includegraphics{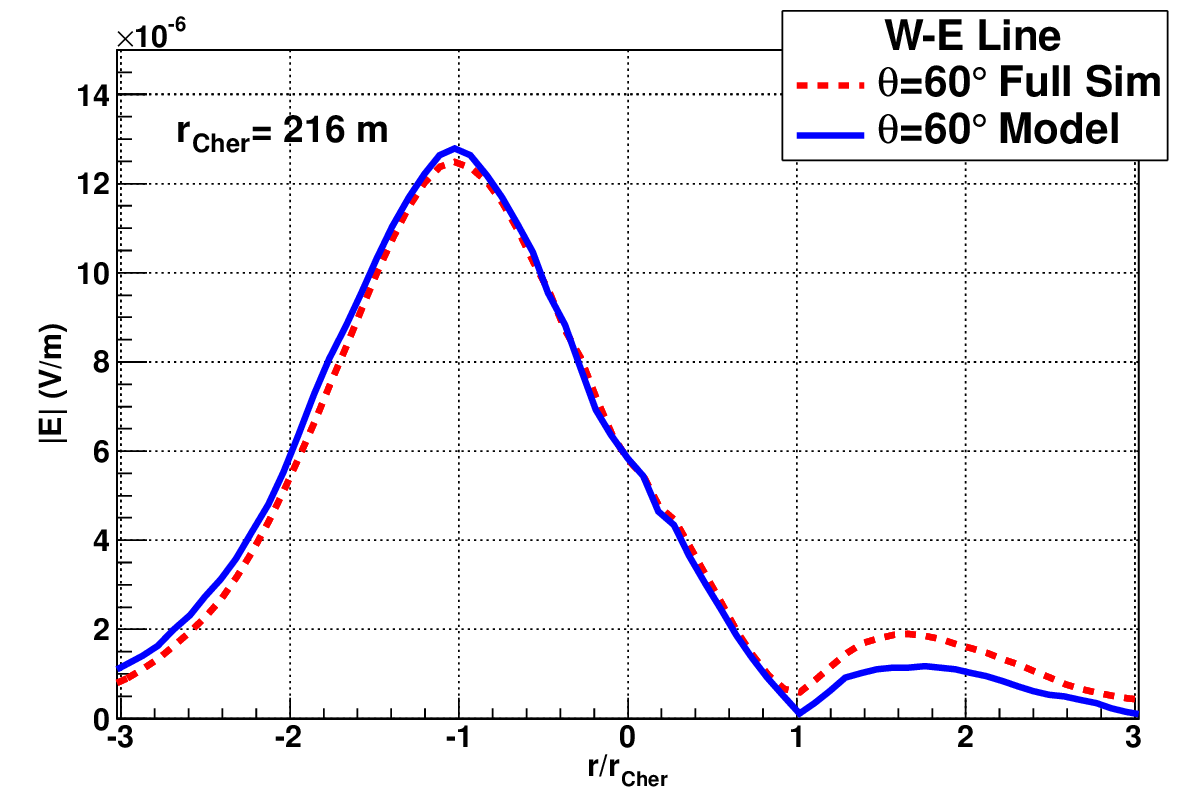}
}
\caption{Comparison of the modulus of the total electric field $\vert
  \vec{{E}}\vert$ predicted by the simple model (solid lines) with that
  obtained in ZHAireS simulations (dashed lines) as a function of distance to
  the core (normalized to the radius of the Cherenkov ring) $r/r_{\rm Cher}$
  for observer positions along the West to East direction (right panels) and
  South to North direction (left panels) through the shower core on
  ground. The geometry corresponds to a proton-induced shower of energy
  $E=10^{17}$ eV coming from the North at $\theta=0^\circ$ (top panels),
  $\theta=30^\circ$ (middle panels), and $\theta=60^\circ$ (bottom panels) at
  the Auger site. 
  The angles between the direction of the shower axis (arriving from North)
  and the magnetic field (at the Auger site) are respectively: 
  $\alpha \sim 53^\circ,~23^\circ~{\rm and}~7^\circ$,
  or equivalently $\sin\alpha \sim 0.80,~0.39~{\rm and}~0.12$ 
  for $\theta=0^\circ,~30^\circ~{\rm and}~60^\circ$.
  Negative values of $r/r_{\rm Cher}$ correspond to the West and South directions in the right and left panels, 
  respectively. The value of $r_{\rm Cher}$ as obtained with Eq.~(\ref{eq:calcring}) is also indicated in the panels.}
\label{fig:badcases}
\end{center}
\end{figure}

It is important to stress that the geometries for which the differences
between model and simulation are more sizeable, represent only a small fraction of the
detectable showers. The reason for this is that the electric field induced in these showers typically has a small 
contribution from the geomagnetic emission mechanism, and as a consequence these showers are less 
likely to trigger an actual array of antennas. Nearly vertical showers develop much
closer to the antennas and can still induce triggers in such situations despite 
the small radio emission, but the solid angle fraction for such showers is small. 
Furthermore, the largest differences between model and
simulation tend to occur in regions where there is a strong destructive
interference between both emission mechanisms, 
therefore the net electric fields are highly suppressed and less likely to be detected.

\section{Discussion and conclusion}
\label{sec:conclusions}

We have shown that by using standard assumptions about the polarization 
and symmetry properties of the electric field induced by the geomagnetic and Askaryan mechanisms, 
we are capable of estimating with great
precision the electric field at any position on the ground,
for a wide range of shower-observer geometries and on an event-by-event basis.
We have shown that in the majority of cases the estimates are within
just a few percent of those obtained by performing full Monte Carlo 
calculations of the electric field from first principles. 
This result represents a quantitative test on the current paradigm under which 
radio emission from atmospheric showers is understood, namely, that the polarized
electric field due to the geomagnetic and charge-excess mechanisms interfere to produce
the complex patterns observed.  
This effort can be seen as an approach that takes into account 
all the complexities of the shower
radio emission in a simple manner, namely by treating the two main emission
mechanisms separately and looking for symmetries that do not apply to the
total net electric field. 

Although not shown in this work, we expect our method to work at other
energies and for other primaries besides the ones used here. We have also discussed  
possible sources of inaccuracies 
as well as features of emission and shower development 
not explicitly taken into account in the model. These increase the differences 
between the predictions and full simulations so that the accuracies reported here 
can be regarded as upper limits on the precision of the model.  
We expect that by refining the assumptions described in Section
\ref{sec:assumptions} we could further improve the accuracy of the model and better 
constrain the two-component paradigm of radio emission in the process.

We expect the proposed method to be very useful in many applications.
It could be used as a test bench for new,
more robust theoretical predictions for the characteristics of each emission mechanism.
Practical applications include the study of
the characteristics of the radio lateral distribution function, such as its
asymmetries and their dependence on shower parameters, and experimental
reconstructions that are based on Monte Carlo simulations of the radio
emission, such as the $X_{\rm max}$ reconstruction of the LOFAR
experiment \cite{LOFAR_Xmax}.  

An obvious advantage of using the method described in this work 
is a significant reduction in computing time. 
As an illustration we note that the two-dimensional
map of the electric field on the ground shown in Fig.~\ref{fig:map}
required only 36 antenna positions East of the core as input\footnote{This number 
of antennas has not been yet optimized and can be significantly further reduced.}. 
The total computing time used
for the input simulations was around 10 h in a single processor.
The input was used to estimate the field at antennas at every 2 m 
in a rectangular grid (totalling around 200 thousand antennas) in a few minutes. 
Our approach could also be a faster alternative to interpolation techniques.

With our model the development and use of extensive parameterizations could be 
largely simplified. Parameterizations are now reduced 
to two simple functions for each shower (the values of $|\vec{E}|$ as a function of distance $R$ 
along the reference line for each of the two mechanisms). Such calculations 
could be used to study more subtle effects such as shower to shower fluctuations or 
those attributed to different hypotheses about composition or hadronic model. 

\section{Acknowledgments}
J.A.M., W.R.C and E.Z. are supported by 
Ministerio de Econom\'\i a (PRI-PRIMASP-2011-1154 and FPA2012-39489),the
Consolider-Ingenio 2010 CPAN Programme (CSD2007-00042) and Xunta de Galicia, Conseller\'\i a de Educaci\'on - Grupos de Referencia
Competitivos (GRC2013-024), Feder Funds, Spain and Marie
Curie-IRSES/EPLANET(European Particle physics Latin American NETwork), European Union 7th Framework Program, Grant No. 
PIRSES-2009-GA-246806. H.S. is supported by Office of Science, U.S. Department of Energy and N.A.S.A.
We thank CESGA (Centro de Supercomputaci\'on de Galicia) for computing resources.




\begin{thebibliography}{99}

\bibitem{Huege_ICRC13}
T. Huege, Proceedings of the $33^{\rm rd}$ ICRC, Rio de Janeiro, Brazil (2013),
available in arXiv:1310.6927 [astro-ph] (2013).
%
\bibitem{Nagano-Watson}
M.~Nagano, A.~Watson, 
Rev. Mod. Phys. {\bf 72} (2000) 689,
and references therein.
%
\bibitem{AERA}
F.G. Schr\"oder for the Pierre Auger Collaboration,
Proceedings of the $33^{\rm rd}$ ICRC, Rio de Janeiro, Brazil (2013), \#0899.
%
\bibitem{EASIER}
R. Gaior for the Pierre Auger Collaboration, 
Proceedings of the $33^{\rm rd}$ ICRC, Rio de Janeiro, Brazil (2013), \#0883
%
\bibitem{Auger}
J. Abraham et al [Auger Collaboration], NIMA, {\bf 523} (2004) 50.
%
\bibitem{CODALEMA}
D. Ardouin {\it et al.} [CODALEMA Collaboration], 
Astropart. Phys. {\bf 31} (2009) 192.
%
\bibitem{LOPES}
A. Nigl {\it et al.} [LOPES Collaboration],
Astron. $\&$ Astrophys. {\bf 488}, 807 (2008).
%
\bibitem{TREND}
D. Ardouin {\it et al.} [TREND Collaboration]
Astropart. Phys. {\bf 34} 717 (2011) 717;
O. Martineau-Huynh for the TREND Collaboration,
arXiv:1204.1599 [astro-ph].
%
\bibitem{LOFAR}
P. Schellart {\it et al.} [LOFAR Collaboration],
Astron. \& Astrophys. {\bf 560} (2013) A98.
%
\bibitem{Tunka-Rex_ICRC13}
F.G. Schr\"oder for the Tunka-Rex Collaboration,
Proceedings of the $33^{\rm rd}$ ICRC, Rio de Janeiro, Brazil (2013), \#0452.
%
\bibitem{LOPES_muons_radio}
W.D. Apel {\it et al.} [KASCADE \& LOPES Collaborations]
Phys. Rev. D {\bf 85} (2012) 071101(R).
%
\bibitem{LOFAR_Xmax}
S. Buitink {\it et al.} [LOFAR Collaboration],
Proceedings of the $33^{\rm rd}$ ICRC, Rio de Janeiro, Brazil (2013), \#0579.
%
\bibitem{Allan_Nature}
H.R. Allan, R.W. Clay, J.K. Jones,
Nature {\bf 227} (1970) 1116.
%
\bibitem{LOPES_LDF}
W.D. Apel {\it et al.} [KASCADE \& LOPES Collaborations]
Astropart. Phys. {\bf 32} (2010) 294.
%
\bibitem{ANITA_UHECR}
S.~Hoover {\it et al.} [ANITA Collaboration],
Phys. Rev. Lett. {\bf 105} (2010) 151101.
%
\bibitem{CROME}
R. Smida {\it et al.} [CROME group],
arXiv:1306.6738 [astro-ph] (2013).
%
\bibitem{Scholten_MGMR}
O. Scholten, K. Werner, F. Rusydi,
Astropart. Phys. {\bf 29} (2008) 94;
K. Werner, O. Scholten,
Astropart. Phys. {\bf 29} (2008) 393.
%
\bibitem{Scholten_EVA}
K. Werner, K. D. de Vries, O. Scholten,
Astropart. Phys. {\bf 37} (2012) 5
%
\bibitem{REAS3}
M. Ludwig, T. Huege,
Astropart. Phys. {\bf 34}, 438 (2011).
%
\bibitem{SELFAS}
V. Marin and B. Revenu,
Astropart. Phys. {\bf 35} (2012) 733.
%
\bibitem{ZHAireS}
 J. Alvarez-Mu\~niz, W.R. Carvalho, E. Zas, 
Astropart. Phys. {\bf 35} (2012) 325.
%
\bibitem{COREAS}
T. Huege, C.W. James, 
Proceedings of the $33^{\rm rd}$ ICRC 2013, Rio de Janeiro, Brazil (2013) \#548. 
%
\bibitem{ZHAireS_ANITA}
 J. Alvarez-Mu\~niz, W.R. Carvalho, A. Romero-Wolf, M. Tueros and E. Zas,
 Phys. Rev. D {\bf 86} (2012) 123007
%
\bibitem{Washington_icrc2013}
J. Alvarez-Mu˜niz, W.R. Carvalho Jr., A. Romero-Wolf, M. Tueros and E. Zas,
Proceedings of the $33^{\rm rd}$ ICRC 2013, Rio de Janeiro, Brazil (2013) \#105.
%
\bibitem{Huege_ARENA12}
T. Huege, AIP Conf. Proc. {\bf 1535} (2013) 121,
avaiable in 1301.2135 [astro-ph] 
%
\bibitem{LOPES_COREAS}
W.D. Apel {\it et al.} [LOPES Collaboration]
Astropart. Phys. {\bf 50-52} (2013) 76
%
\bibitem{AERA_pol}
A. Aab {\it et al.} [Pierre Auger Collaboration]
Phys. Rev. D {\bf 89}, 052002 (2014).
%
\bibitem{Huege_radio_mass}
T. Huege, R. Ulrich, R. Engel, 
Astropart. Phys. {\bf 30} (2008) 96.
%
\bibitem{Konstantinov_JCAP}
N.N. Kalmykov and A.A. Konstantinov
JCAP {\bf 12} (2012) 029
%
\bibitem{deVries_radio_mass}
K.D. de Vries, A.M. van den Berg, O. Scholten, K. Werner, 
Astropart. Phys. {\bf 34} (2010) 267.
%
\bibitem{Scholten_Xmax}
K. D. de Vries, O. Scholten, K. Werner,
Astropart. Phys. {\bf 45} (2013) 23–27.
%
\bibitem{LOFAR_param_paper}
A. Nelles {\it et al.}
arXiv:1402.2872  [astro-ph]
%
\bibitem{Allan}
H. R. Allan, in: J. G. Wilson, S. A. Wouthuysen (Eds.), Progress in
Elementary Particle and Cosmic Ray Physics {\bf 10}, North Holland, (1971), p. 169,
and refs. therein.
%
\bibitem{Prescott}
J.R. Prescott, J.H. Hough, and J.K. Pidcock, 
Nature Phys. Science 233 (1971) 109.
%
\bibitem{Askaryan} G.A. Askar'yan, 
Soviet Physics JETP {\bf 14} (1962) 441; 
G. A. Askaryan, Soviet Phys. JETP {\bf 21} (1965) 658.
%
\bibitem{Kahn-Lerche}
F.D.~Kahn, I.~Lerche,
Proc. Roy. Soc. A{\bf 289} (1966) 206.
%
%
\bibitem{Jackson}
J.D. Jackson, 
``Classical Electrodynamics" $3^{\rm rd}$ Ed.  (Wiley, New York, 1998).
%
\bibitem{ZHS92} E. Zas, F. Halzen, T. Stanev, 
Phys. Rev. D {\bf 45} (1992) 362.
%
\bibitem{ZHS91} F.~Halzen, E.~Zas, T.~Stanev, 
Phys. Lett. B {\bf 257} (1991) 432. 
%
\bibitem{ZHS13} 
D. Garc\'\i a-Fern\'andez, J.~Alvarez-Mu\~niz, W.R. Carvalho Jr., A. Romero-Wolf, E. Zas, 
Phys. Rev. D {\bf 87} (2013) 023003.
%
\bibitem{Scholten_PRL}
K. D. de Vries, A. M. van den Berg, O. Scholten, K. Werner 
Phys. Rev. Lett. {\bf 107} (2011) 061101

\end{thebibliography}
\end{document}